# Derivation of a governing rule in triboelectric charging and series from thermoelectricity

**Authors:** Eui-Cheol Shin[1], Jae-Hyeon Ko[2], Ho-Ki Lyeo[3], Yong-Hyun Kim[1,2]*

**Affiliations:**

[1]Department of Physics, Korea Advanced Institute of Science and Technology (KAIST), Daejeon 34141, Republic of Korea.

[2]Graduate School of Nanoscience and Technology, KAIST, Daejeon 34141, Republic of Korea.

[3]Korea Research Institute of Standards and Science, Daejeon 34113, Republic of Korea.

*Correspondence to: yong.hyun.kim@kaist.ac.kr

**Abstract:** Friction-driven static electrification is familiar and fundamental in daily life, industry, and technology, but its basics have long been unknown and have continually perplexed scientists from ancient Greece to the modern high-tech era. Despite its simple manifestation, triboelectric charging is believed to be very complex because of the unresolvable interfacial interaction between two rubbing materials. Here, we for the first time reveal a simple physics of triboelectric charging and triboelectric series based on friction-originated thermoelectric charging effects at the interface, characterized by the material density ($\rho$), specific heat ($c$), thermal conductivity ($k$), and Seebeck coefficient ($S$) of each material. We demonstrate that energy dissipational heat at the interface induces temperature variations in the materials and thus develops electrostatic potentials that will initiate thermoelectric charging across the interface. We find that the trends and quantities of triboelectric charging for various polymers, metals, semiconductors, and even lightning clouds are simply governed by the triboelectric factor $\xi = S/\sqrt{\rho c k}$. The triboelectric figure-of-merit is expressed with the triboelectric power $K = \xi\sqrt{t/\pi}$, of which the difference can be maximized up to 1.2 V/W·cm$^{-2}$ at the friction time $t = 1$ s. Our findings will bring significant opportunities for microscopic understanding and management of triboelectricity or static electrification.

# Ⅰ. INTRODUCTION

Triboelectric charging is mysterious because its fundamental origin has not been known for a long time [1-13]. The phenomenon itself is obvious, as in combing hair or lightning in daily life, and humanity discovered electricity from it and came to understand almost every aspect of electricity and magnetism through Maxwell's equations. But why do we not yet rigorously know which material is charged positively or negatively when two materials are rubbed in spite of the great success of quantum mechanics and condensed matter physics? What is the fundamental origin of the mysteriousness of triboelectric charging?

Here is a collection of even more puzzling facts about triboelectric charging, gleaned after thousands of years of observations.

(1) Triboelectric charging occurs universally. It occurs not only at solid-solid contacts, but also at various solid-liquid [14], solid-gas [15], liquid-liquid [16], and gas-gas [17] systems. Lateral friction dominantly induces a charging effect called triboelectrification [13], but normal contact also causes electrification to a large extent, in a process called contact electrification [5-8].

(2) Generally, triboelectric charging effects are negligibly small for metallic systems, but maximized for insulating polymeric systems [18].

(3) Triboelectric series do exist. Surprisingly, there is no consensus on the triboelectric series accepted by the community [19]. A general tendency was observed in the experiment, but reproducibility is not settled at the level of science. Middle-school textbooks have started to remove the triboelectric series from the contents [20], simply because it is not accurate enough.

(4) Even more peculiar triboelectric effects have been routinely observed; identical materials exhibit charging effects when rubbed together [1,21-23]; dust particles are charged depending on their size [24,25]; and groups of materials exhibit cyclic triboelectric charging [18,26].

To date, no single theory can satisfactorily explain this mysterious but fundamental phenomenon. The failure has been reluctantly attributed to the unknown complexity of interfacial interaction between two contacting materials [2,13]. Many researchers have proposed various scenarios for triboelectric charging, such as electron transfer due to work-function difference [3,4], direct ion or material transfer [6,7], thermionic emission [8], or tribo-emission [9], mechanochemistry [13], and flexoelectricity [10]. Each theory may explain a specific case or more, but its general application is very limited. With no general guiding principle, recent developments of the triboelectric-based energy harvesting technology [8,11,12] will be fundamentally hindered. Also, static electrification, which sometimes leads to fires at gas stations, lightning during air operations, and unintentional damage to electronic devices, is a growing worry for modern industry [27] and has to be controlled, if possible, via microscopic management.

Our goal here is to derive a simple governing rule of the mysterious triboelectric charging. To do this, we start with the assumption that a successful triboelectric theory should meet two conditions: (A) the triboelectric charging effect should follow a causality rule as other physics laws do; (B) the causality rule should be universally applicable. For example, the electron transfer due to work-function difference may explain contact electrification between metals, but this could happen without friction, violating condition (A). Thermionic emission could explain electron transfer when two solid materials contact at a strongly-repulsive region, but this cannot happen at a gas-gas contact, violating condition (B).

The cause of the triboelectric charging effect is definitely friction between two contacting materials. Friction must be explicitly or implicitly connected to the electrification to satisfy condition (A). To make the cause more general to satisfy condition (B), we view the friction as *energy dissipation* regardless of its detailed complexity at the interface. This particularly makes sense at the first order for light frictional situations without severe damage or material transfer. Then, we know that the energy dissipation or frictional heat conduction at the interface between two materials results in temperature variations in the materials and an abrupt temperature drop at the interface [28]. We could naturally expect thermoelectric charge-redistribution effects according to the temperature profile [29-31]. Can this friction-originated thermoelectric charge redistribution be a solution to the mystery of triboelectric charging? At least, this scenario meets two conditions (A) and (B), as the thermoelectric effect is generally applicable in gases, liquids, and solids.

At the end, following the friction-originated thermoelectric charging scenario, we are able to rigorously quantify the triboelectric series, as shown in Fig. 1(a), by the triboelectric factor $\xi = S/\sqrt{\rho c k}$, which is expressed by the thermodynamic material properties such as the material density ($\rho$), specific heat ($c$), thermal conductivity ($k$), and Seebeck coefficient ($S$) of each material. Remarkably, the triboelectric series has an absolute origin at $S = 0$, and each material can be triboelectrically positive or negative depending on its Seebeck coefficient $S$. Based on our theory, the Seebeck coefficient $S$ is the fundamental source of the mysteriousness of triboelectricity. Now we will demonstrate how the thermoelectric charging scenario leads to the unprecedented quantification of the triboelectric series.

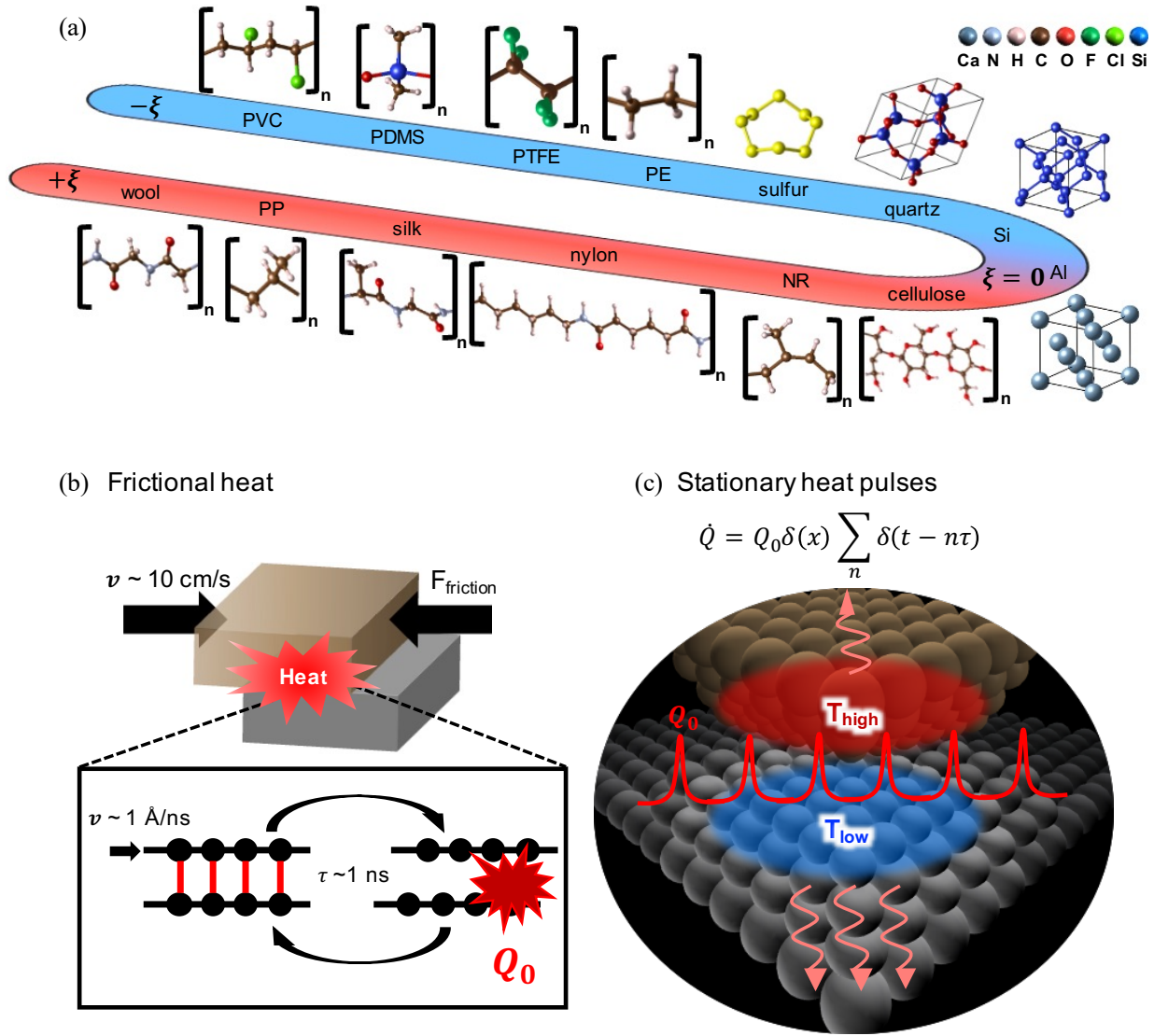

FIG. 1. Triboelectric series and frictional heat. (a) Triboelectric series quantified with triboelectric factor $\xi$ for various triboelectric materials. The polymeric unit or crystal structure is also displayed. The triboelectric series has its origin at $\xi = 0$ for metals or superconductors for which the Seebeck coefficient is zero. (b) Frictional heat generated at the interface of two contacting materials. Microscopically, the frictional heat is associated with the bond-breaking process due to external mechanical motion. We assume that the relative speed of the two materials is 10 cm/s or 1 Å/ns,

and the bond-breaking process repeats every $\tau = 1$ ns. (c) Stationary heat pulses $\dot{Q}(x,t) = Q_0 \delta(x) \sum_n \delta(t - n\tau)$ located at the interface, representing frictional heat from successive bond-breaking processes. The temperature profile near the interface can be different with $T_{high}$ and $T_{low}$, and there is an abrupt temperature drop across the interface depending on thermal properties of materials.

## II. FRICTIONAL HEAT GENERATION AND THERMOELECTRIC EFFECTS

When two materials are rubbed together, the mechanical energy dissipates into frictional heat. Microscopically, frictional heat is generated via a bond-breaking process at the interface, as schematically shown in Fig. 1(b). If the relative speed of the two contacting materials is about $v = 10$ cm/s, or 1 Å/ns, approximately $Q_0 = 0.01$ J/m² of heat is generated at the interface per $\tau = 1$ ns (see Appendix A for estimation). To make the problem simple, we replace the frictional motion with a stationary heat source $\dot{Q}(x,t) = Q_0 \delta(x) \sum_n^N \delta(t - n\tau)$ at the interface that generates successive $N$ heat pulses, as shown in Fig. 1(c). Here, $\tau$ is the period of heat pulses, and $t$ is the time for which friction applies. The $\delta$-function-like heat pulse is a reasonable assumption because bond-breaking and phonon excitation typically occur very quickly, on the time scale of sub-picoseconds [32-34]. Also, the stationary heat generator model is naturally applicable to multiple contact electrification, in which similar bond-breaking energy dissipation occurs at the interface during the vertical contact-separation process but with a much larger time period $\tau$ of ms to seconds.

Generally, heat flows away from a heat source, and heat conduction is governed by Fourier's law $\dot{Q} = -k\nabla T$. Here, $\dot{Q}$ is the heat density flux and $\nabla T$ is the temperature gradient. Once a temperature profile $T$ is settled in a longer timescale, a thermo-electromotive force instantaneously develops due to the thermoelectric effect [31,35] according to $J = \sigma(E - S\nabla T)$, where $J$ is the current density, $\sigma$ the electrical conductivity, and $E$ the electric field. At the steady-state, the open-circuit condition $J = 0$ is applied throughout the materials. Then, the electrostatic potential $V$ is simply described by

$$V = -ST \quad (1)$$

since $E = -\nabla V$, and the accordingly redistributed charge density $\rho_e$ is expressed as $\rho_e = \varepsilon S \nabla^2 T$ from Gauss' law, where $\varepsilon$ is the dielectric constant. Therefore, if we can determine the temperature profile $T$ from the heat conduction equation, we can obtain the electrostatic potential profile $V$ and the charge density profile $\rho_e$ with information of $\varepsilon$ and $S$ [31]. In this way, any thermal electrification between two contacting materials can be evaluated.

In a thermoelectric experiment with two contacting materials [31,36], heat flows from one material to the other. While temperature gradually varies within the materials, it suddenly drops at the interface. The abrupt interfacial temperature drop [28] is known to be determined by the interfacial thermal conductance $\kappa$. In our triboelectric model, the heat source is located at the interface of two materials, as schematically shown in Fig. S1 of Supplemental Material (SM) [37]. Therefore, the interfacial temperature will be high, with a certain temperature difference between the two materials, and the temperatures will gradually decrease to the ambient temperature $T_0$. Figure S1 [37] also shows trivial linear solutions of temperature profiles of finite-size materials for thermoelectricity and triboelectricity at the steady-state.

### III. SEEBECK COEFFICIENT FROM QUANTUM MECHANICS

To proceed further, we need information about the Seebeck coefficient for various triboelectric materials. Unfortunately, Seebeck coefficients for triboelectric materials have not been well documented in experiments. This is partly because most triboelectric materials are not generally good conductors to measure Seebeck coefficients, or simply because we have not been interested in the property in this regard. In theory [31,38], however, the Seebeck coefficient is easily calculable from first-principles quantum mechanical calculations and sensitively depends on the location of the Fermi energy $E_F$ and the local density of states at $E_F$. The location of $E_F$ of a material is determined by various sample qualities and environment conditions and can readily vary even for a single material; for example, a fraction of charge will cause a wide swing of $E_F$ for wide-gap insulating materials such as polymers. Because the $E_F$ position for specific materials is also not known from experiments, we have to devise a way to locate $E_F$ while considering general triboelectric environments. Our *ad hoc* choice is the universal alignment of the Fermi energy among various semiconductor materials and water [39]. As moisture is inevitable in the air to some extent, we take the $H_2/H^+$ redox potential at $-4.44$ eV as the most probable location of $E_F$ for various triboelectric materials.

With this choice, we quantum mechanically calculate the band alignment and Seebeck coefficient of representative triboelectric materials in Fig. 1(a) – wool (sulfur-crosslinked α-keratin), polypropylene (PP), silk, nylon, polyisoprene (NR: natural rubber), cellulose, Al, Si, quartz ($SiO_2$), sulfur, polyethylene (PE), polytetrafluoroethylene (PTFE), polydimethylsiloxane (PDMS: silicone rubber), and polyvinyl chloride (PVC) [40-47] – using first-principles density-functional theory (DFT) formulation, as shown in Fig. 2 (see Appendix B for computational details). When the charge neutrality position $E_{\text{neutral}}$ of a material is higher than the global Fermi energy $E_F$ at $-4.44$ eV, the Seebeck coefficient becomes positive. Metallic Al has an almost zero Seebeck coefficient. Finally, when $E_F$ is located above the $E_{\text{neutral}}$ of a material, the Seebeck coefficient becomes negative. Generally, the magnitude of the Seebeck coefficient increases as the separation between $E_F$ and $E_{\text{neutral}}$ decreases [38].

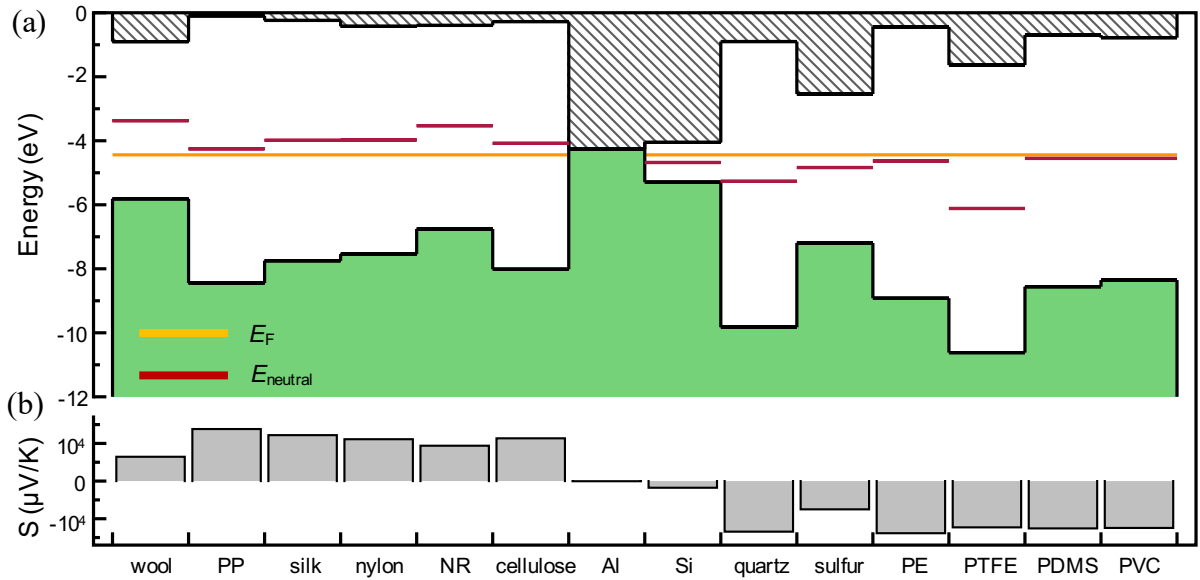

FIG. 2. Electronic structure and Seebeck coefficient of triboelectric materials. (a) Electronic structure of representative triboelectric materials (see text for full names) with the valence (filled) and conduction (half-filled) bands from DFT calculations. The band offsets are aligned with reference to the vacuum level at zero energy, and the Fermi energy $E_F$ is universally aligned with the hydrogen redox level at $-4.44$ eV. The charge-neutrality level for each material is marked as the neutral energy $E_{\text{neutral}}$. (b) DFT-calculated Seebeck coefficient $S$ at assumed $E_F$ for triboelectric materials.

In reality, the Fermi energy is not perfectly aligned and shows non-negligible variations depending on materials [39]. Also, the crystal structures that we used in DFT calculations do not represent real materials 100%. Therefore, we admit that our current Seebeck coefficients may contain appreciable errors in magnitude or sign. One relief is that the electronic structures of weakly-interacting polymeric materials show no big changes depending on detailed crystal structures from first-principles DFT calculations [48]. With some tolerance in mind, it is still worthwhile to discuss, as a proof of concepts, thermoelectric charging and its tendency based on these *ad hoc* values, before any accurately-measured value is available.

## IV. HEAT PARTITION AND INTERFACIAL THERMAL CONDUCTANCE

To obtain the temperature profile of two *semi-infinite* materials in contact with an interfacial heat generator $\dot{Q}(x,t)$, we have to solve the one-dimensional heat equation, derived from Fourier's law [35],

$$\frac{\partial T}{\partial t} = \frac{k}{\rho c}\frac{\partial^2 T}{\partial x^2} + \frac{\dot{Q}}{\rho c} \qquad (2)$$

where $\alpha = k/\rho c$ is called thermal diffusivity. This equation is numerically solvable, but an analytical solution also exists with the help of two simplifications below.

*Heat partition*: When heat is generated from a bond-breaking process at an interface, the issue of how much the initial heat is partitioned into the two contacting materials 1 and 2 is non-trivial. As $Q = \rho c l T$, the initial heat needs to be partitioned depending on the volume (or the depth $l$) and $T = T_{\text{flash}} - T_0$, where $T_{\text{flash}}$ is the instantaneously increased temperature [49]. We can think of two postulates of four different cases, as shown in Fig. 3, to determine $T_{1,2}$ that are necessary for solving the heat equation in Eq. (2). One postulate is that the flash temperatures are the same, $T_1 = T_2$, but with different depth $l$, *i.e.*, (i) the first atomic layer, (ii) the phonon mean-free-path, or (iii) the thermal diffusion length of each material. The other postulate is that (iv) the heat is just partitioned by half to the depth of the phonon mean-free-path and the flash temperatures are not the same, $T_1 \neq T_2$. We have found that only the latter postulate (iv) provides separable triboelectric characteristics for each material (see Appendix C for details). The half-and-half postulate (iv) will be mainly used in our discussion for simplicity, but the other same-temperature postulate will provide similar physics but with a more complicated coupled formula, as displayed in Table S1 [37].

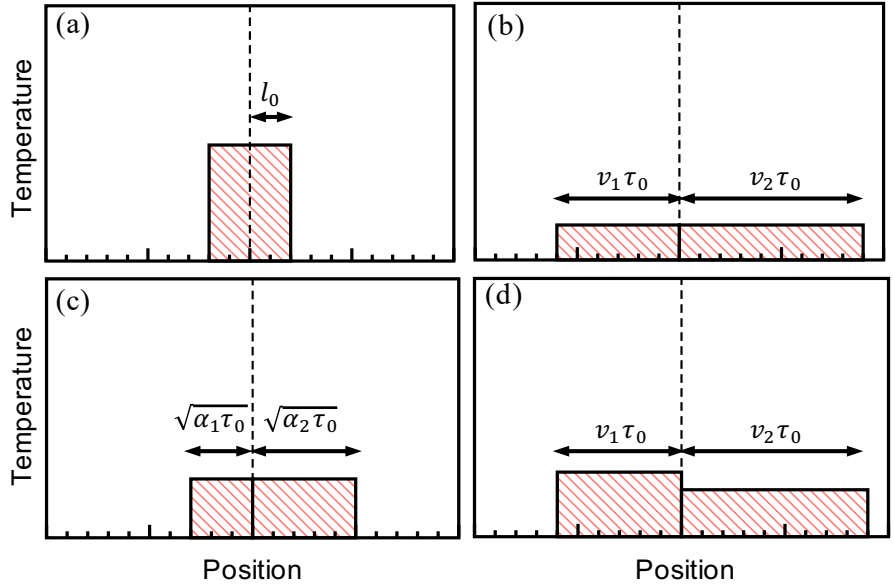

FIG. 3. Heat partition. Initial heat produced at the interface can be partitioned into two materials following four different schemes. (a) Partitioned heat is distributed within the first atomic layer $l_0$ with the same flash temperature. (b) Partitioned heat is distributed within the ballistic phonon mean-free-path $v\tau_0$ with the same flash temperature. $v$ is the sound velocity and $\tau_0$ is the short heating time of less than 0.1 ps. (c) Partitioned heat is distributed within the thermal diffusion length $\sqrt{\alpha\tau_0}$ with the same flash temperature. (d) Initial heat is partitioned by a half and distributed within the ballistic phonon mean-free-path $v\tau_0$. The flash temperatures are not the same.

*Interfacial thermal conductance*: Depending on the interfacial thermal conductance $\kappa$, heat exchange occurs between two contacting materials, resulting in a change of the temperature difference $\Delta T_{\text{int}} = T_1 - T_2$ at the interface. When $\kappa$ is large (over 10 MW/m²·K), $\Delta T_{\text{int}}$ is noticeably affected. But when $\kappa$ is in the range of 0.001-0.1 MW/m²·K, the variation of $\Delta T_{\text{int}}$ is negligible, as shown in Fig. 4. So, we set $\kappa = 0$ as a good approximation for triboelectric charging from weak interfacial couplings.

With these simplifications, we can separately and analytically obtain temperature profiles of individual materials.

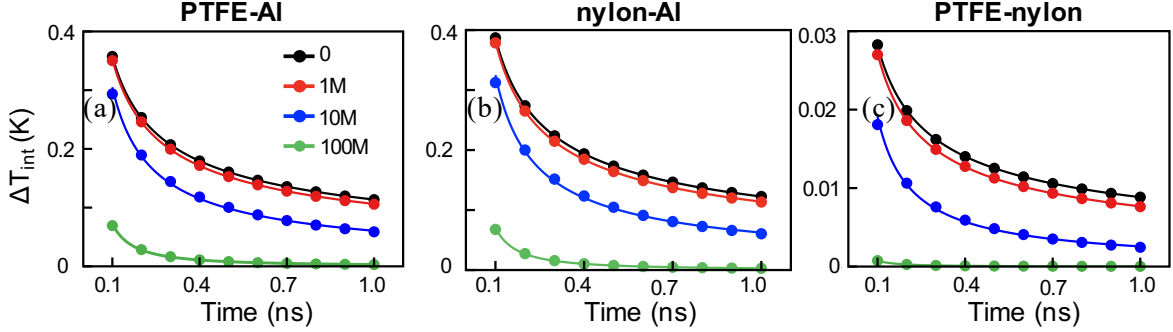

FIG. 4. Interfacial thermal conductance. (a)-(c) Evolution of temperature gap $\Delta T_{\text{int}}$ at various values of interfacial thermal conductance $\kappa$ from 0 to 100 MW/m²·K, from numerical solutions for three representative triboelectric pairs (PTFE-Al, nylon-Al, and PTFE-nylon). When $\kappa$ is less than 1 MW/m²·K in the weak-coupling regime, the change of $\Delta T_{\text{int}}$ is insignificant, justifying the assumption $\kappa = 0$ used for analytic solutions.

## V. TEMPERATURE PROFILES AND TRIBOELECTRIC CHARGING

To obtain temperature profiles, we solve Eq. (2) for three pairs of three prominent triboelectric materials; PTFE is known to be triboelectrically negative, and its $S < 0$; nylon is known to be positive and its $S > 0$; metallic Al is amphoteric and its $S$ is nearly zero. We first numerically calculate temperature profiles while considering the non-zero interfacial thermal conductance $\kappa$ and using experimental values for $\rho$, $c$, and $k$ for three materials (see Appendix D for details). We then obtain the analytical solution using Green's functions when $\kappa = 0$,

$$T(x,t) = \frac{1}{2} \frac{Q_0}{\sqrt{\pi \rho c k}} \sum_{n=0}^{N} \frac{e^{-\frac{x^2}{4\alpha(t-n\tau)}}}{\sqrt{t-n\tau}} =_{t \gg \tau} \frac{1}{2} \frac{\frac{Q_0}{\tau}}{\sqrt{\pi \rho c k}} \sqrt{t}\, \mathcal{E}_{\frac{3}{2}}\left[\frac{x^2}{4\alpha t}\right] \quad (3)$$

where $x$ is the position from the interface (see Appendix E for the detailed derivation). The summation should be done up to $N < t/\tau$. When the summation is converted into the integral for large $t \gg \tau$, the special integral function $\mathcal{E}_m(z) = \int_1^\infty e^{-zu}/u^m du$ appears in the final form. Figure S2 [37] confirms that numerical solutions when $\kappa = 0$ match exactly with the analytical solutions.

Figure 5(a)-(c) shows the evolution of the temperature profiles near the interface for three triboelectric pairs (PTFE-Al, nylon-Al, and PTFE-nylon) for a single heat pulse from Eq. (3) before the second pulse arrives. For a single heat pulse at the interface, the temperature profile exhibits a Gaussian shape with the dispersion of $2\sqrt{\alpha t}$. Due to their difference in thermal conductivity, PTFE and nylon stay at a relatively high temperature at the interface, whereas metallic Al cools down quickly. Because of this, we see noticeable temperature gaps for PTFE-Al and nylon-Al pairs and a small gap for the polymer pair. In our numerical solutions for all three pairs, as shown in Fig. 4, the temperature gap $\Delta T_{\text{int}}$ is reduced slightly (~10%) when the interfacial thermal conductance $\kappa$ is 1 MW/m$^2$·K.

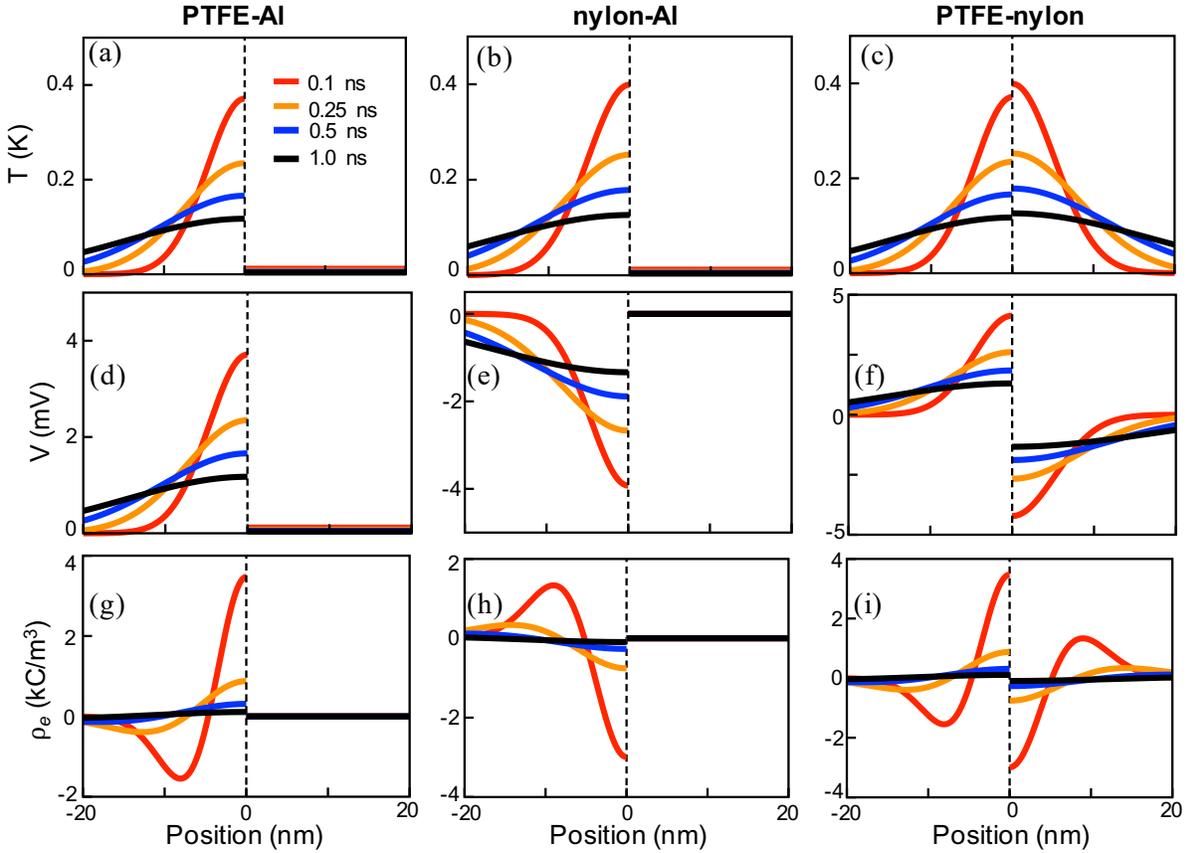

FIG. 5. Triboelectricity for a single heat pulse. (a)-(c) Evolution of temperature profiles after a single heat pulse at the interface for three representative triboelectric pairs (PTFE-Al, nylon-Al,

and PTFE-nylon). The analytical solution in Eq. (3) is used for the plot. An abrupt temperature gap occurs at the interface. (d)-(f) Evolution of voltage profiles after a single heat pulse at the interface. We simply used Eq. (1) by multiplying $-S$ to $T(x,t)$. (g)-(i) Evolution of charge density profiles for the three triboelectric pairs. The charge density profile was obtained using $\rho_e = \varepsilon S \nabla^2 T$ with theoretical Seebeck coefficient $S$ and dielectric constant $\varepsilon$.

With the temperature profiles and the Seebeck coefficients, we calculate the electrostatic potential profiles from Eq. (1) and the charge density distributions from $\rho_e = \varepsilon S \nabla^2 T$, as displayed in Fig. 5(d)-(f) and Fig. 5(g)-(i), respectively. We see that the polarity of $S$ determines the polarity of $V$ and $\rho_e$. Note that the open-circuit condition $J = 0$ is used in our derivation. This means that we do not allow any charge transfer across the interface, even though there is a potential difference between two contacting materials. In this adiabatic condition, the net charge of each material should be zero. When we allow charge transfer by shorting two materials at the interface, as happens in reality, a net charge will move across the interface following the electrostatic potential difference. For the PTFE-Al pair in Fig. 5(d), a net charge will move from high-potential PTFE to low-potential Al, completing the triboelectric charging effect; PTFE becomes negative and Al becomes positive. On the other hand, Al will be negatively charged for the nylon-Al pair. This reproduces the general experimental triboelectric tendency of the PTFE-Al-nylon series. The PTFE-nylon pair will show greater charge transfer because of the larger potential difference, in spite of the small temperature gap.

When more heat pulses in Eq. (3) are considered for a longer time, the temperature profile becomes non-Gaussian in shape (see Fig. S3 [37]). Figure 6(a)-(c) shows that, when $N$ or $t$ increases, the temperature profile spreads to $\mu$m with gradually increasing $T$ at the interface. Figure 6(d)-(f) and Fig. 6(g)-(i) show corresponding voltage profiles and charge density distributions, respectively, for the three triboelectric pairs for large $N$ up to 1000, corresponding to 1 $\mu$s. As $t$ increases, charge is accumulated in the range of ~30 nm from the interface and grows like a $\delta$-function. The additional charge accumulation or depletion for triboelectric insulators can be understood in terms of band bending effects due to the thermo-electromotive force $S\nabla T$ along the temperature gradient $\nabla T$.

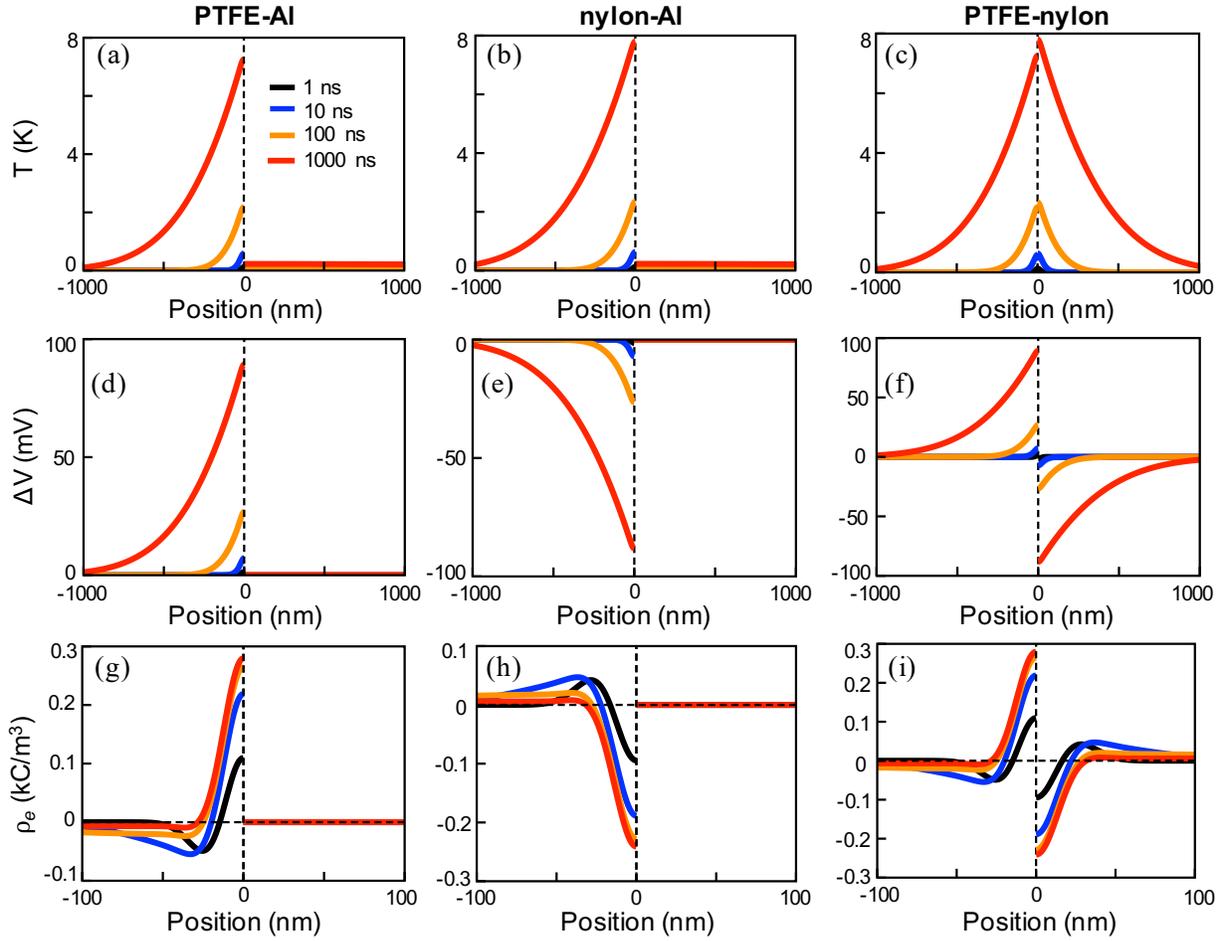

FIG. 6. Time evolution of triboelectricity. (a)-(c) Evolution of temperature profiles for multiple heat pulses up to 1000 for three representative triboelectric pairs (PTFE-Al, nylon-Al, and PTFE-nylon). (d)-(f) Evolution of electrostatic potential or voltage profiles $\Delta V$ at the interface for the three triboelectric pairs. The voltage profile quickly spreads to the depth of 1 $\mu$m within a few $\mu$s, and the interfacial voltage difference becomes appreciable. (g)-(i) Evolution of charge density profiles $\rho_e$ up to 1 $\mu$s for the three triboelectric pairs. Localized positive charge accumulates within 30 nm of the interface for PTFE as time increases; negative charge accumulates for nylon. Note that the current flow is not allowed to cross the interface because of the open-circuit condition $J = 0$ that we used.

If we integrate the same charge near the interface for various $t$, the surface charge converges to a constant value after several ms, as shown in Fig. 7. Ultimately, the constant surface charge density is evaluated as

$$\sigma_{\text{surface}} = -\frac{\varepsilon S}{2k}\frac{Q_0}{\tau} \tag{4}$$

which is localized at the interface (see Appendix F for the detailed derivation). The $\delta$-accumulated surface charge $\sigma_{\text{surface}}$ is a kind of reservoir for triboelectric charging, ready to move across the interface when the short-circuit is allowed. The monotonic increase of surface charge density is corroborated to a certain extent by experimental observation of surface charge increase in multiple-contact electrifications [23,50]. The magnitude of $\sigma_{\text{surface}}$ is comparable to the experimental values [50] of ~3 μC/m² although it is an adiabatic quantity with the opposite polarity.

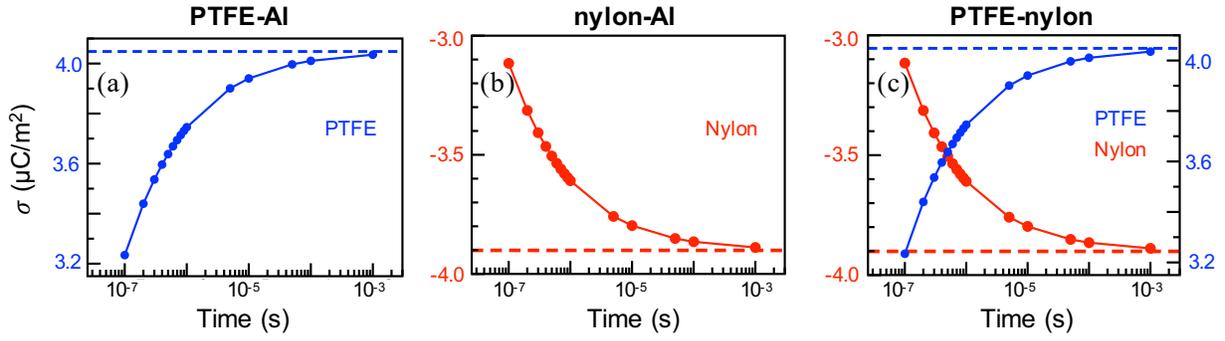

FIG. 7. Constant surface charge density. (a)-(c) Evolution of surface charge densities up to 1 ms for three representative triboelectric pairs (PTFE-Al, nylon-Al, and PTFE-nylon). The surface charge density $\sigma_{\text{surface}}$ is obtained by integrating only the localized interfacial charge with the same polarity in Fig. 6(g)-(i). $\sigma_{\text{surface}}$ converges to a constant value. Al's surface charge density is negligible (not shown).

## VI. QUANTITATIVE TRIBOELECTRIC SERIES

Now we realize that triboelectric charging is solely determined by the potential difference at the interface $x = 0$ of two contacting materials, i.e., $V_1 - V_2 = S_2 T_2(0,t) - S_1 T_1(0,t)$. Note that this equation is typically used for evaluating a thermocouple [35] that measures the open-circuit voltage of a hot spot (see Fig. S4 [37]). Indeed, our triboelectric model is equivalent to a thermocouple. Because $\mathcal{E}_m(0) = 1/(m-1)$ when $m > 1$, we obtain the potential difference from Eq. (3) as

$$V_1 - V_2 = \frac{Q_0}{\tau}\sqrt{t/\pi}\left[\frac{S_2}{\sqrt{\rho_2 c_2 k_2}} - \frac{S_1}{\sqrt{\rho_1 c_1 k_1}}\right]. \tag{5}$$

This indicates that triboelectric charging is, remarkably, determined by the material characteristics $S, \rho, c$, and $k$ of each material. We also find that the potential difference or triboelectric charging increases with the $\sqrt{t}$-dependency as the friction application time $t$ increases.

Considering the polarity of triboelectric charging after the short-circuit condition in Eq. (5), we define a triboelectric factor $\xi$ that rigorously quantifies the triboelectric series as

$$\xi = \frac{S}{\sqrt{\rho c k}} \tag{6}$$

where the Seebeck coefficient $S$ determines the absolute zero and sign of the triboelectric factor of a material. We suggest that the triboelectric series based on the triboelectric factor $\xi$ is more fundamental than the transferred charge based quantification in Eq. (4) or experiments [11,12]. In other words, triboelectric series should be defined by the direction, but not by the amount of charge transfer due to friction.

Following Eq. (6), we display the triboelectric series of the representative materials in Fig. 8(a) using theoretical $S$ and experimental $\rho, c$, and $k$ values (see Appendix G). Note that $\xi$ is large for small $\rho$ and $k$, e.g., as in polymeric materials. It may be possible to obtain more accurate triboelectric series for well-characterized material samples, particularly with carefully-measured values of $S$ and $k$. One interesting thing about the new quantity $\xi$ is its unit V·s$^{1/2}$/J·cm$^{-2}$, which may be associated with electrostatic potential per unit energy density, but not exactly because of s$^{1/2}$.

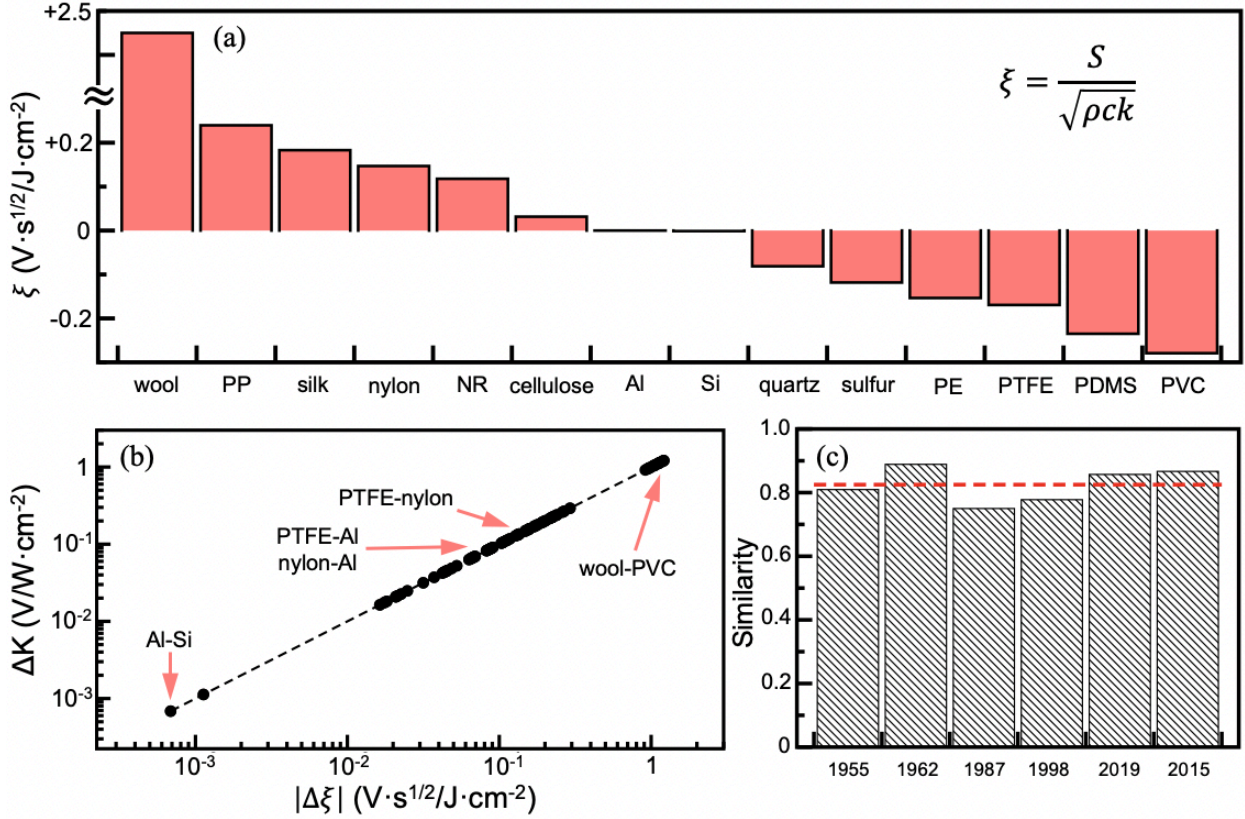

FIG. 8. Quantitative triboelectric series and figure-of-merit. (a) Triboelectric series quantified by triboelectric factor $\xi = S/\sqrt{\rho c k}$. The Seebeck coefficient determines the sign and zero of the triboelectric series. Polymeric materials have large $\xi$ values because of large $S$ and low $\rho$ (density) and $k$ (thermal conductivity). (b) Triboelectric figure-of-merit quantified by the triboelectric power difference $\Delta K = \Delta \xi \sqrt{t/\pi}$ for various triboelectric pairs at $t = 1$ s. The wool-PVC pair shows the largest $\Delta K$ of 1.2 V/W·cm$^{-2}$. (c) The similarity of various triboelectric series in the literature (Table I) with respect to the quantitative triboelectric series of the present study. The average similarity is about 83%.

The charge is not yet transferred in our current derivation of triboelectric charging. Therefore, it is awkward to define the efficiency of triboelectricity. Still, we can use Eq. (5) to evaluate the strength of triboelectric charging. We propose triboelectric power $K$ as a new figure-of-merit of triboelectricity,

$$K = \xi \sqrt{t/\pi} \qquad (7)$$

of which the unit is V/W·cm$^{-2}$, representing the electrostatic potential per unit power density. Then, we obtain triboelectric voltage $\Delta V = \Delta K \frac{Q_0}{\tau}$, which is analogous to the thermoelectric formula $\Delta V = S\Delta T$ with the thermoelectric power $S$ in the unit of V/K, representing the electrostatic potential per unit temperature. The triboelectric power $K$ becomes more significant with the $\sqrt{t}$-dependency when the friction time $t$ increases. In Fig. 8(b), we plot the triboelectric power difference $\Delta K$ for various triboelectric pairs for $t = 1$ s, reaching 1.2 V/W·cm$^{-2}$ for the wool-PVC pair.

Table I. Triboelectric series reported in the literature [11,12,20,51-54] and the present study for representative triboelectric materials including wool, polypropylene (PP), silk, nylon, polyisoprene (NR: natural rubber), cellulose, Al, Si, quartz (c-SiO$_2$), sulfur, polyethylene (PE), polytetrafluoroethylene (PTFE), polydimethylsiloxane (PDMS: silicone rubber), and polyvinyl chloride (PVC). Despite some ambiguity in the experimental materials, we consider fur, hair, and wool as sulfur-crosslinked α-keratin, cotton and wood as cellulose, and rubber as silicone rubber. The triboelectric series in the middle school text [20] has been removed in the recent edition.

| **Most positive** ──────────────────────────▶ **Most negative** | Pub. year |
|---|---|
| wool → *nylon* → *cotton* → silk → PVC → PE → **PTFE** | 1955[51] |
| *nylon* → wool → silk → *cotton* → NR → S → PE → PVC → **PTFE** | 1962[52] |
| *nylon* → wool → silk → paper → *cotton* → PE → PP → PVC → Si → **PTFE** | 1987[53] |
| quartz → *nylon* → wool → silk → *cotton* → paper → metals → rubber → **PTFE** → PVC | 1998[54] |
| Copy paper → *nylon* → PP → quartz → PE → PDMS → **PTFE** → PVC | 2019[11,12] |
| fur → glass → silk → *wood* → rubber → plastic | 2015[20] |
| wool → PP → silk → *nylon* → NR → *cellulose* → Al → Si → quartz → S → PE → **PTFE** → PDMS → PVC | 2022[present] |

To check the reliability of our quantitative triboelectric series, we compare similarity between various triboelectric series (see Appendix H for details), as shown in Fig. 8(c). We found that our quantitative triboelectric series is on average about 83% similar to other experimental triboelectric series reported in the literature [11,12,20,51-54], as listed in Table I. We can clearly recognize a general order of nylon, cellulose, and PTFE. While the order of wool, silk, and nylon is not settled in the series, we know that their $\xi$-values can be sensitively modified with detailed values of $S$, $\rho$, $c$, and $k$. The remarkable similarity reflects that, in spite of various levels of approximations in our triboelectric theory, the thermoelectric charging scenario quite reasonably captures the essence of what happens in triboelectricity. Now we cautiously claim that the unsolved mystery of triboelectric charging can be attributed to the sensitive sample-by-sample dependency of the $\xi$-value or $S$. This may make sense particularly for polymeric insulators, for which the Fermi energy and thus $S$ can be sensitively altered by sample quality and environment.

**VII. LIGHTNING CLOUDS**

To demonstrate the general applicability of our heat-based triboelectric theory to non-solid systems, we apply the theory to the problem of triboelectric charging in lightning clouds. The physical quantities $S$, $\rho$, $c$, and $k$ could be well defined in theory for gases and liquids, although the thermoelectricity of gases and liquids might be more complicated [55]. Assuming a rigid contact between two clouds, we can roughly estimate the behavior of triboelectric charging in lightning clouds, as shown in Figs. 9(a) and (b). For simplicity, we treat the air as an ideal gas, following $pV = \left(\frac{m}{M}\right)RT$ or $p = \rho RT/M$, where $p$ is the pressure, $V$ the volume, $m$ the mass, $M$ the mass of 1-mole air, and $R$ the gas constant. We also assume that two airs have different base temperatures $T_1$ and $T_2$, but nearly the same pressure $p$ and the same Seebeck coefficient $S$. The thermal conductivity of an ideal gas has a temperature dependency of $k \sim \sqrt{T}$. As $\rho c$ is proportional to $1/T$, $\rho c k$ is proportional to $1/\sqrt{T}$. Then, the triboelectric power difference of the two frictional airs can be written as

$$\Delta K = K_1 - K_2 = \frac{S\sqrt{t/\pi}}{\sqrt{\rho_1 c_1 k_1}} \left[1 - \left(\frac{T_2}{T_1}\right)^{\frac{1}{4}}\right] \quad (8)$$

which indicates that $\Delta K$ is positive if $T_2/T_1$ is less than 1. The triboelectric factor $\xi = S/\sqrt{\rho c k}$ of air is 20.4 V·s$^{1/2}$/J·cm$^{-2}$, which is much larger than those of typical triboelectric materials in Fig.

8(a). Once the triboelectric potential difference is developed following Eq. (8), various charge carriers may directly move across the interface of lightning clouds. Figures 9(a) and 9(c) show that the rising higher-temperature air is triboelectrically positively charged, and the lower-temperature air is negatively charged [17].

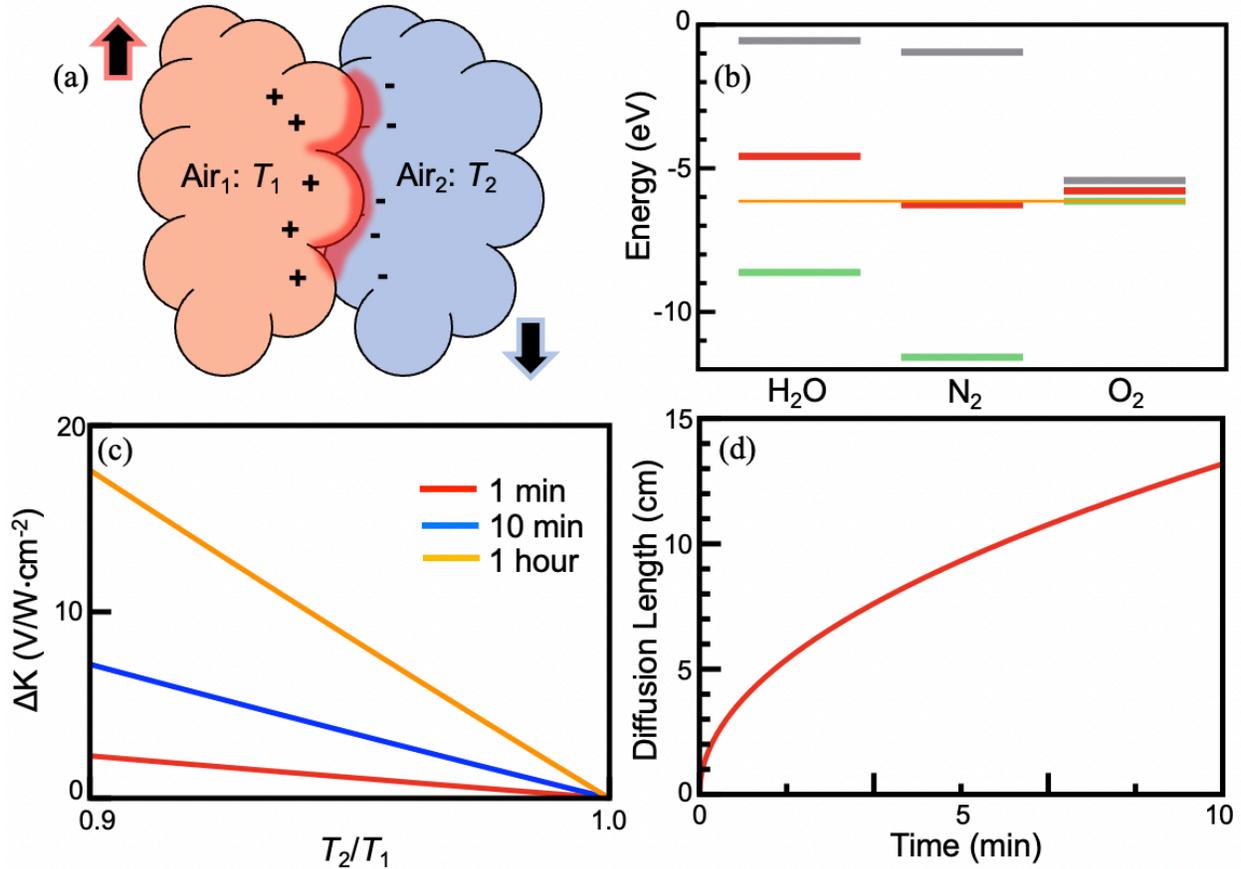

FIG. 9. Triboelectricity in lightning clouds. (a) Schematic of two frictional airs with different temperatures $T_1$ and $T_2$ in lightning clouds. We found that the higher-temperature rising air is positively charged, and the lower-temperature sinking air is negatively charged. (b) DFT-aligned molecular orbitals (MO) of $H_2O$, $N_2$, and $O_2$ in air. The highest occupied molecular orbital (HOMO) and the lowest unoccupied molecular orbital (LUMO) are marked by green and grey solid lines, respectively. The Fermi energy and the charge neutrality position are marked by orange and red solid lines, respectively. Because $O_2$ takes electrons easily from the environment, we set the Fermi energy to the HOMO of $O_2$. The Seebeck coefficient of air was calculated to be 0.01 V/K for $H_2O$ molecules. (c) Triboelectric power difference $\Delta K = K_1 - K_2$ of two frictional airs depending on the ratio of temperatures at $t = 1$ min, 10 min, and 1 hour. The positive value indicates that the higher-temperature air is charged positively. (d) Thermal diffusion length $2\sqrt{\alpha t}$ of the interfacial temperature profile of lightning airs as a function of frictional time $t$. For this estimation, we used $\rho = 1.25$ kg/m³, $c = 736$ J/kg·K, and $k = 0.026$ W/m·K [56]. The exponential temperature distribution spreads up to 13 cm at $t = 10$ min.

**VIII. DISCUSSION**

The triboelectric nanogenerator (TENG) [8,11,12] has been around for about a decade as a plausible energy harvesting technology from various mechanical energy sources. The technology is believed to be very close to commercialization. Although there have been many engineering attempts to improve its device design for better efficiency, there was no specific guiding rule for material selection. Our quantitative theory clearly suggests how to select triboelectric materials and how to control individual properties to achieve high triboelectric figure-of-merits. The theory can also be used for minimizing static electrification and preventing lightning damage.

New quantities for triboelectric charging, *i.e.*, the triboelectric factor $\xi$ and the triboelectric power $K$ in Eqs. (6) and (7), can serve as a golden rule for understanding and microscopic management of static electrification. For example, to maximize triboelectric charging for energy harvesting, one needs to design sparse or porous materials with large Seebeck coefficients, small material densities $\rho$, and small thermal conductivities $k$. Generally, organic polymer materials possess these properties. It is also desirable to maximize $\Delta\xi$ such that the Seebeck coefficient can be induced to have opposite polarity by intentional or unintentional doping. This may be readily achievable in inorganic semiconductor materials. To this end, organic-inorganic hybrid materials such as colloidal quantum-dot solids could be a good candidate. Finally, the unique power-law dependency of $\sqrt{t}$ implies that a longer friction time is limitedly better for energy harvesting.

**IX. CONCLUSION**

We rigorously proved that simple thermoelectric physics exists in triboelectric charging and the triboelectric series, which have been unsolved for a long time. Friction-originated thermoelectric charging is a key to quantitatively understanding the fundamental phenomenon.

After the unprecedented quantification $\xi = S/\sqrt{\rho c k}$ of triboelectric series materials, we can now see why static electrification occurs significantly for polymeric insulating materials and negligibly for metallic materials. Our findings will pave the way to solving various mysteries of triboelectric charging and offer significant opportunities for controlling static electrification at the microscopic level.

**Acknowledgments:** We thank H.-H. Nahm, J. Y. Park, S.-Y. Kim, S. Jeong, and J. Kang for discussion. This work was supported by National Research Foundation of Korea (NRF) grants (2016R1A5A1008184, 2019R1A6A1A10073887, 2019M3D1A1078302) and by the Grand Challenge 30 program from the College of Natural Sciences, KAIST.

**Author contributions:** Y.-H.K. conceived the idea and designed the research. E.-C.S. and Y.-H.K. developed the theory, performed the calculations, and wrote the manuscript. J.-H.K. contributed to the early stage of the research. H.-K.L. contributed to the interpretation of the research. All authors commented on the manuscript.

**Competing interests:** Y.-H.K. and E.-C.S. have a patent application (KR-10-2021-0000860, 10-2021-0049278, pending in the Republic of Korea), which disclose methods for designing and evaluating triboelectric materials and energy harvesters.

**Data and materials availability:** All data is available in the main text or the supplementary materials, or through the corresponding author upon reasonable request.

**APPENDIX A: ROUGH ESTIMATION OF FRICTIONAL HEAT**

We assume a relative speed $v$ of 10 cm/s or 1 Å/ns for the two rubbing materials. The typical magnitude of van der Waals (vdW) interactions $E_{\text{vdw}}$ is on the order of 1 KJ/mol [57]. We can think of an energy density of about $E_{\text{vdw}}/A$ for a one-to-one vdW bond of an ideal square lattice with a lattice constant of 1 Å, where $A$ is the contact area defined by 1 Å$^2$. As a result, vdW energy between the two materials is approximately 10 meV/Å$^2$, as shown in Fig. S5 [37]. This estimated value can be justified by first-principles DFT calculations. Table S2 [37] shows the vdW binding energy $E_{\text{b}}$ per contact area for several systems (Fig. S6 [37]). For both the polymer-polymer and the metal-polymer contacts, we obtain an order of 10 meV/Å$^2$, consistent with the above estimation for the ideal square lattice. Then, mechanical energy is supplied to break the vdW bonding, and only a fraction of the bond-breaking energy may be converted into the frictional heat $Q$. The fraction may be associated with the friction coefficient $\mu$, which is assumed to be 0.1. Then, the heat generation rate is roughly estimated to be $Q_0 = 0.01$ J/m$^2$ per $\tau = 1$ ns. We used this value for numerical estimations. Frictional heat can be generated continuously or in the form of pulses.

**APPENDIX B: DFT CALCULATION OF SEEBECK COEFFICIENTS**

To obtain theoretical Seebeck coefficients, we performed first-principles DFT calculations using the Vienna *ab initio* simulation program (VASP) with projector-augmented wave (PAW) pseudopotentials in the VASP database [58] and the Perdew-Burke-Ernzerhof (PBE) functional [59] for various triboelectric materials, for which crystal structures are known from experiments, as displayed in Figs. S7-S10 [37]. To correct the underestimated DFT band gaps, we used the Heyd-Scuseria-Ernzerhof (HSE06) exchange-correlation functional [60] with mixing parameters from the reciprocal of the macroscopic dielectric constant $1/\varepsilon_{\text{PBE}}^{\infty}$ [61] and a screening parameter of 0.2 Å$^{-1}$. The mixing parameters based on $\varepsilon_{\text{PBE}}^{\infty}$ provide the significantly reduced error for large band gap systems such as polymeric materials. The macroscopic dielectric constant was calculated within the random phase approximation scheme using PBE wavefunctions, as listed in Table S3 [37]. This dielectric constant was also used for calculating the charge-density distribution. Band alignment was done using multilayer slab models and their vacuum potentials defined by the Hartree potential [62], which is commonly shared in PBE and HSE06 calculations.

The kinetic energy cutoff of 500 eV was used for all calculations. All materials were relaxed until the Hellman-Feynman atomic forces were less than 0.03 eV/Å. The linear tetrahedron method was used for the Brillouin-zone sampling with the following grid: (4×2×2) for $\alpha$-keratin; (4×1×4) for polypropylene (PP); (4×4×4) for silk (fibroin); (2×4×4) for nylon; (4×2×2) for polyisoprene (NR); (4×2×2) for cellulose; (12×12×12) for Al; (4×4×4) for Si; (4×4×4) for quartz (SiO$_2$); (4×4×2) for sulfur; (4×2×2) for polyethylene (PE); (4×2×2) for polytetrafluoroethylene (PTFE); (4×2×4) for polydimethylsiloxane (PDMS) and polyvinyl chloride (PVC).

The Seebeck coefficient was calculated [38] at the universally-aligned Fermi energy $E_F$ using

$$S(E_F) = -\frac{1}{eT_0} \frac{\int N(E)(E - E_F)\left(-\frac{df}{dE}\right) dE}{\int N(E)\left(-\frac{df}{dE}\right) dE} \quad (B1)$$

where $e$ is the electron charge, $T_0$ the ambient temperature (300 K), $N(E)$ the DFT density of states of each material, and $f$ the Fermi-Dirac distribution function. In general, $S$ is positive if $E_F$ is lower than the charge-neutrality energy $E_{\text{neutral}}$, and vice versa. $S = 0$ if $E_F = E_{\text{neutral}}$.

It may be crucial to check whether theoretical Seebeck coefficients from Eq. (B1) properly represent experimental values at least in orders of magnitude. Therefore, we compared the experimentally well-documented Seebeck coefficients of Al and Si with DFT-based theoretical values. The experimental Seebeck coefficient of Al is -1.8 µV/K [63], while the calculated one is -2.5 µV/K at 300 K. For Si, the experimental Seebeck coefficient is -673 µV/K [64], very close to the calculated one (-880 µV/K) for n-type Si with a carrier density of $2 \times 10^{18}$ cm$^{-3}$ at 325 K.

**APPENDIX C: HEAT PARTITION MODELS**

The heat partition in Table S1 [37] and the initial temperature profile $T_{\text{ini}}$ in Fig. 3 were derived using the following equations;

(i) First atomic layer: $Q_0 = Q_1 + Q_2 = \rho_1 c_1 l_0 T_{\text{ini}} + \rho_2 c_2 l_0 T_{\text{ini}}$

(ii) Phonon mean free path: $Q_0 = Q_1 + Q_2 = \rho_1 c_1 v_1 \tau_0 T_{\text{ini}} + \rho_2 c_2 v_2 \tau_0 T_{\text{ini}}$

(iii) Thermal diffusion length: $Q_0 = Q_1 + Q_2 = \rho_1 c_1 \sqrt{\alpha_1 \tau_0}\, T_{\text{ini}} + \rho_2 c_2 \sqrt{\alpha_2 \tau_0}\, T_{\text{ini}}$

(iv) Half and half: $Q_1 = Q_0/2 = \rho_1 c_1 v_1 \tau_0 T_{\text{ini}}^1$; $Q_2 = Q_0/2 = \rho_2 c_2 v_2 \tau_0 T_{\text{ini}}^2$

where $\tau_0$ is the short heating time.

Using the initial temperature profile $T_{\text{ini}}$, we were able to obtain the evolution of the temperature profile using Green's function [65], as follows,

$$T(x, t) = \frac{1}{\sqrt{4\pi\alpha t}} \int_0^\infty \left( e^{-\frac{(x+y)^2}{4\alpha t}} + e^{-\frac{(x-y)^2}{4\alpha t}} \right) T_{\text{ini}}(y)\, dy \quad (C1)$$

where $T_{\text{ini}}(y)$ is the initial temperature profile. Because we assumed the square function of $T_{\text{ini}}(y)$, the integral becomes

$$T(x,t) = \frac{T_{\text{ini}}}{\sqrt{4\pi\alpha t}} \int_0^l \left( e^{-\frac{(x+y)^2}{4\alpha t}} + e^{-\frac{(x-y)^2}{4\alpha t}} \right) dy = \frac{T_{\text{ini}}}{2} \left[ \text{Erf}\left(\frac{x+l}{2\sqrt{\alpha t}}\right) - \text{Erf}\left(\frac{x-l}{2\sqrt{\alpha t}}\right) \right] \quad (C2)$$

where the error function is used. When $l$ is very small at $\tau_0 \to 0$, we obtained the following temperature profiles:

(i) First atomic layer:

$$T_{1,2}(x,t) = \frac{Q_0}{\rho_1 c_1 + \rho_2 c_2} \frac{e^{-\frac{x^2}{4\alpha_{1,2}t}}}{\sqrt{\pi\alpha_{1,2}t}} \quad (C3)$$

(ii) Phonon mean free path:

$$T_{1,2}(x,t) = \frac{Q_0 v_{1,2}}{\rho_1 c_1 v_1 + \rho_2 c_2 v_2} \frac{e^{-\frac{x^2}{4\alpha_{1,2}t}}}{\sqrt{\pi\alpha_{1,2}t}} \quad (C4)$$

(iii) Thermal diffusion length:

$$T_{1,2}(x,t) = \frac{Q_0}{\sqrt{\rho_1 c_1 k_1} + \sqrt{\rho_2 c_2 k_2}} \frac{e^{-\frac{x^2}{4\alpha_{1,2}t}}}{\sqrt{\pi t}} \quad (C5)$$

(iv) Half and half:

$$T_{1,2}(x,t) = \frac{Q_0}{2} \frac{e^{-\frac{x^2}{4\alpha_{1,2}t}}}{\sqrt{\rho_{1,2} c_{1,2} k_{1,2} \pi t}} \quad (C6)$$

Note that the half-and-half temperature profile is completely separable and has no dependency on the sound velocity $v$ in the final form.

**APPENDIX D: NUMERICAL SOLUTIONS FOR TEMPERATURE PROFILES**

Using the backward-Euler method [66], the one-dimensional heat equation in Eq. (2) was numerically solved for two semi-infinite materials in contact. Different from the explicit Euler method, the implicit method gives stable, non-oscillatory solutions for the heat equation in our numerical simulations. The system size was large enough for the propagating heat not to reach the other boundary whose temperature was kept at ambient temperature $T_0$. The initial heat $Q_0$ is injected instantaneously within 0.1 ps and propagates with a time step of 1 ps. The interfacial thermal conductance $\kappa$ is fully considered to allow heat exchange at the interface.

**APPENDIX E: ANALYTICAL SOLUTIONS FOR TEMPERATURE PROFILES**

Because we set $\kappa = 0$, we solved Eq. (2) analytically for individual semi-infinite materials using Green's function [65], as follows,

$$T(x,t) = \int_0^s \int_0^\infty \frac{\left(e^{-\frac{(x-y)^2}{4\alpha(t-s)}} + e^{-\frac{(x-y)^2}{4\alpha(t-s)}}\right)}{\sqrt{4\pi\alpha(t-s)}} \frac{\dot{Q}(y,s)}{\rho c} \, dy \, ds \quad (E1)$$

with periodic heat pulses $\dot{Q}$ at the surface,

$$\dot{Q}(x,t) = \frac{1}{2} Q_0 \delta(x) \sum_{n=0}^{N} \delta(t - n\tau) \quad (E2)$$

where 1/2 comes from the heat partition postulate (see Table S1 [37]). Time $t$ should be larger than $N\tau$. Then, the spatial and time $\delta$-function integrations give

$$T(x,t) = \frac{Q_0}{2\rho c} \sum_{n=0}^{N} \int_0^s \frac{e^{-\frac{x^2}{4\alpha(t-s)}}}{\sqrt{4\pi\alpha(t-s)}} \delta(s - n\tau) \, ds \quad (E3)$$

$$= \frac{Q_0}{2\sqrt{\pi\rho c k}} \sum_{n=0}^{N} \frac{e^{-\frac{x^2}{4\alpha(t-n\tau)}}}{\sqrt{t - n\tau}} \quad (E4)$$

which can be converted into an integral when $t \gg \tau$. In our estimation, we mostly used $t = 1$ s and $\tau = 1$ ns, satisfying the condition $t \gg \tau$. Because we know that

$$\sum_{n=0}^{N} \frac{e^{-\frac{x^2}{4\alpha(t-n\tau)}}}{\sqrt{t - n\tau}} = \int_0^{N\tau} \frac{e^{-\frac{x^2}{4\alpha(t-z)}}}{\sqrt{t - z}} \frac{dz}{\tau}, \quad (E5)$$

we derive the analytic temperature profile as

$$T(x,t) = \frac{\frac{Q_0}{\tau}}{2\sqrt{\pi\rho ck}} \left[ \sqrt{t}\, \mathcal{E}_{\frac{3}{2}}\left[\frac{x^2}{4\alpha t}\right] - \sqrt{t-N\tau}\, \mathcal{E}_{\frac{3}{2}}\left[\frac{x^2}{4\alpha(t-N\tau)}\right] \right] \tag{E6}$$

which consists of two terms, $T_{\text{first}}$ and $T_{\text{second}}$, with different dispersion lengths and opposite polarity. The first term $T_{\text{first}}$ is the same as Eq. (3). The second term $T_{\text{second}}$ is negligible compared to the first term $T_{\text{first}}$ for the condition $t \gg \tau$. $\mathcal{E}_n(x)$ is a special function called the general exponential integral, defined as

$$\mathcal{E}_m(z) = \int_1^\infty \frac{e^{-zu}}{u^m} du \tag{E7}$$

which can be evaluated numerically.

The final solution in Eq. (3) is also valid for the continuous injection of heat at the surface with the rate of $Q_0/\tau$, as shown in Fig. S11 [37], justifying our use of the pulse model. From Green's function method [65] of Eq. (3), we obtain the temperature profile as

$$T(x,t) = \frac{1}{2}\frac{\frac{Q_0}{\tau}}{\rho c}\int_0^t \frac{e^{-\frac{x^2}{4\alpha(t-s)}}}{\sqrt{4\pi\alpha(t-s)}} ds \tag{E8}$$

$$= \frac{1}{2}\frac{\frac{Q_0}{\tau}}{\sqrt{\pi\rho ck}} \sqrt{t}\, \mathcal{E}_{\frac{3}{2}}\left[\frac{x^2}{4\alpha t}\right] \tag{E9}$$

which is exactly the same as Eq. (3) for $t \gg \tau$. Physically, this makes sense because both the two heat generation schemes are indistinguishable for large $t \gg \tau$. Our pulse model, however, provides a richer story about the triboelectric charging process, as we discussed in the main text. Also, our stationary heat generator model at the interface for static electrification is applicable not only for frictional heat, but also for other heat sources such as interfacial chemical reactions, artificial heaters, and laser pulses.

## APPENDIX F: SURFACE CHARGE DENSITY

With the information of $\varepsilon$ and $S$, the charge density distribution can be obtained from the second derivative of the temperature profile as $\rho_e = \varepsilon S \nabla^2 T$. Because no charge escaped from the semi-infinite material, the net charge should always be zero, as $\int_0^\infty \rho_e dx = 0$. As Eq. (E6) consists of two terms, $T_{\text{first}}$ and $T_{\text{second}}$, the integral of the charge density distribution also consists of two terms, $\sigma_{\text{first}}$ and $\sigma_{\text{second}}$, which are equal in magnitude but opposite in sign. $\sigma_{\text{first}}$ corresponds to the inner charge because its distribution is very wide with the dispersion length of $2\sqrt{\alpha t}$. On the other hand, $\sigma_{\text{second}}$ is very localized at the interface with the $2\sqrt{\alpha(t-N\tau)}$ dispersion, thus corresponding to the $\delta$-function-like surface-charge density, $\sigma_{\text{surface}}$. Then, $\sigma_{\text{surface}}$ was obtained by integrating $\varepsilon S \nabla^2 T_{\text{second}}$ as

$$\sigma_{\text{surface}} = \int_0^\infty \varepsilon S \nabla^2 T_{\text{second}}(x,t) dx \tag{F1}$$

which results in Eq. (4), which is independent of time $t$.

## APPENDIX G: TRIBOELECTRIC MATERIAL CHARACTERISTICS

Table II. Triboelectric material characteristics are listed. The material density ($\rho$), specific heat ($c$), and thermal conductivity ($k$) are taken from experimental results; $I\beta$-cellulose's are taken from molecular dynamics simulations. The calculated Seebeck coefficient and triboelectric factor are also displayed.

|  | $\rho_{\text{th}}$ (g/cm³) | $\rho$ (g/cm³) | $c$ (J/g·K) | $k$ (W/m·K) | $S$ (μV/K) | $\xi$ (V·s$^{1/2}$/J·cm$^{-2}$) |
|---|---|---|---|---|---|---|
| wool [67,68] | 0.82 | 0.026 | 1.37 | 0.034 | 6458 | 1.86 |
| PP [69] | 0.93 | 0.9 | 1.68 | 0.22 | 13821 | 0.239 |
| silk [70] | 1.24 | 1.4 | 1.24 | 0.256 | 12201 | 0.183 |
| nylon [71,72] | 0.95 | 1.24 | 1.50 | 0.27 | 11118 | 0.156 |

| Material | | | | | | |
|---|---|---|---|---|---|---|
| NR [73,74] | 1.01 | 0.96 | 1.89 | 0.35 | 9392 | 0.117 |
| cellulose [75-77] | 1.35 | 1.6 | 1.40 | 5.7 | 12604 | 0.035 |
| Al [78] | 2.71 | 2.70 | 0.95 | 238 | -2.5 | $-1 \times 10^{-6}$ |
| silicon [79] | 2.28 | 2.33 | 0.70 | 130 | -1769 | $-1.2 \times 10^{-3}$ |
| quartz [78] | 2.49 | 2.5 | 0.78 | 1.4 | -13434 | -0.081 |
| sulfur [78] | 1.47 | 2.07 | 0.72 | 0.27 | -7492 | -0.118 |
| PE [80] | 0.80 | 0.93 | 1.83 | 0.46 | -13846 | -0.156 |
| PTFE [81] | 2.27 | 2.20 | 1.05 | 0.26 | -12291 | -0.158 |
| PDMS [72] | 2.25 | 0.97 | 1.6 | 0.2 | -12568 | -0.225 |
| PVC [78] | 1.44 | 1.38 | 0.96 | 0.15 | -12449 | -0.279 |

**APPENDIX H: SIMILARITY OF TRIBOELECTRIC SERIES**

To compare the two triboelectric series, we define a similarity function of $s(M_i, M_j) = 1$ if the triboelectric order for $M_i$ and $M_j$ materials in a series is the same with respect to the reference series, and $s(M_i, M_j) = 0$ if the order is different. We chose our quantitative triboelectric series as the reference. Average $\bar{s}$ was calculated for all possible combinations. If two series had the same order, then $\bar{s} = 1$. If two series were random, then $\bar{s} = 0.5$. If two series were completely opposite, $\bar{s} = 0$. Averaged value of $\bar{s}$ for five experimental series with reference to the quantitative series was about 0.83.

# Supplemental Material for

# Derivation of a governing rule in triboelectric charging and series from thermoelectricity

Eui-Cheol Shin[1], Jae-Hyeon Ko[2], Ho-Ki Lyeo[3], Yong-Hyun Kim[1,2]*

[1]Department of Physics, Korea Advanced Institute of Science and Technology (KAIST), Daejeon 34141, Republic of Korea.

[2]Graduate School of Nanoscience and Technology, KAIST, Daejeon 34141, Republic of Korea.

[3]Korea Research Institute of Standards and Science, Daejeon 34113, Republic of Korea.

*Correspondence to: yong.hyun.kim@kaist.ac.kr

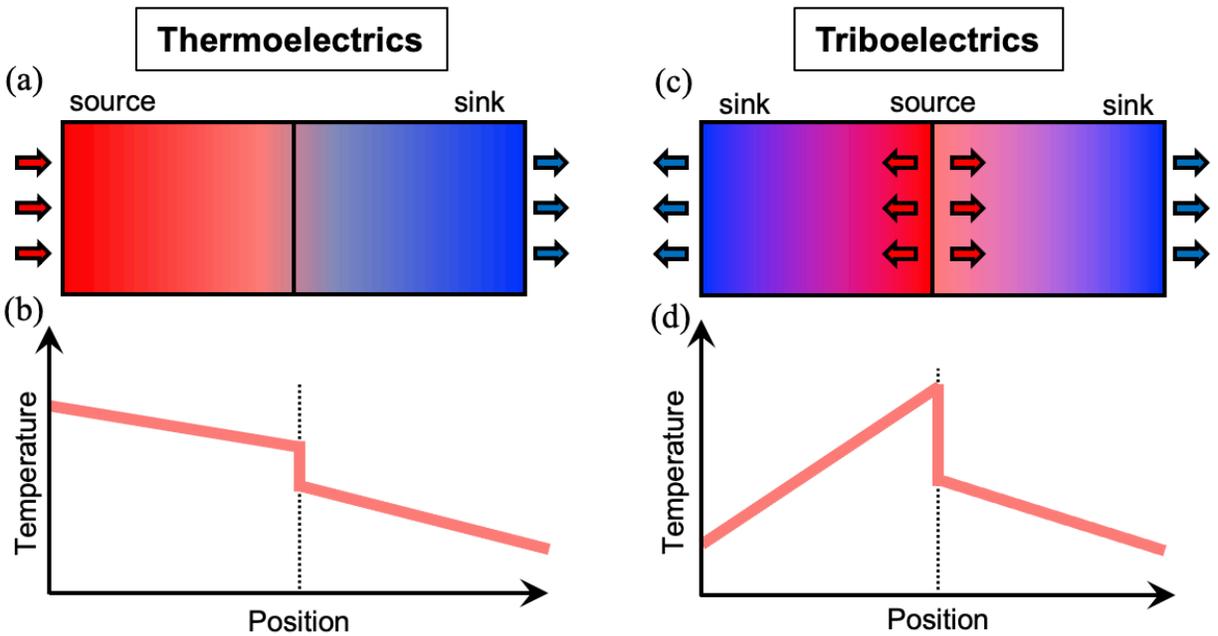

FIG. S1. Thermoelectricity *vs.* triboelectricity. (a)-(b) Schematic of heat transfer from one side to the other side of two in-contact finite-size materials in thermoelectric experiments and the corresponding temperature profile. At the steady state, temperature decreases linearly within the materials depending on thermal conductivity and drops abruptly at the interface depending on the interfacial thermal conductance. (c)-(d) Schematic of heat transfer from an interfacial heat source to both sides of two finite-size materials in triboelectric experiments and the corresponding temperature profile, which is a trivial solution of the heat equation at the steady state. A temperature gap is expected at the interface.

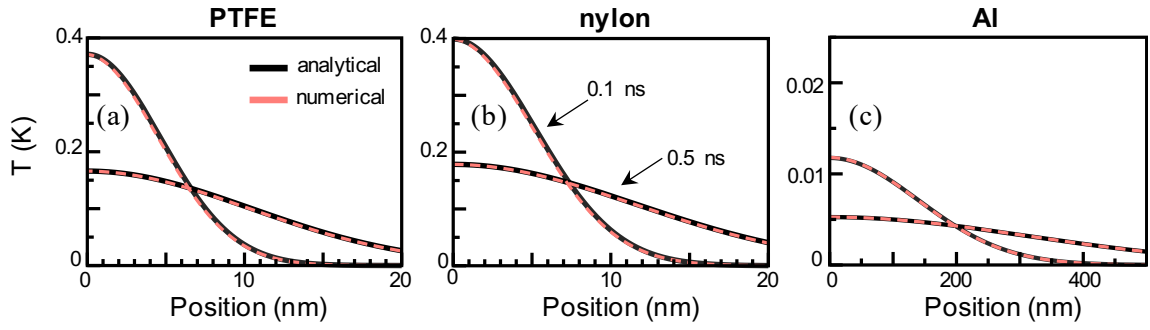

FIG. S2. Numerical *vs.* analytical solutions. Evolution of temperature profiles for a single heat pulse from numerical (pink) and analytical (black) solutions for (a) PTFE, (b) nylon, and (c) Al.

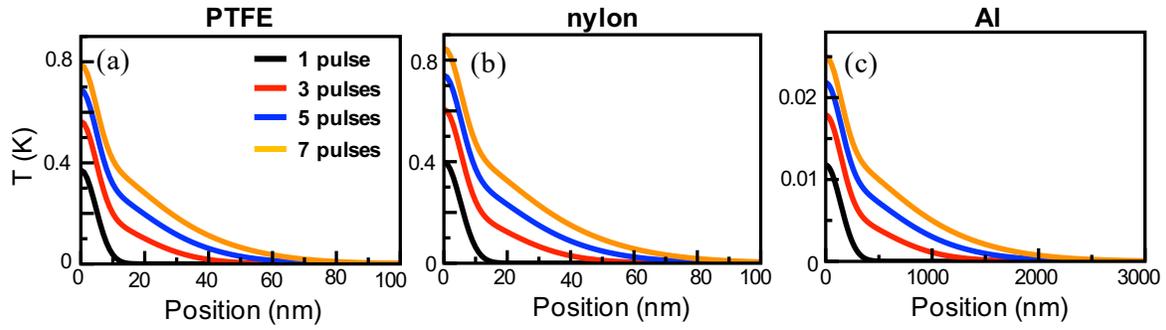

FIG. S3. Temperature profiles for multiple heat pulses. Evolution of temperature profiles after multiple heat pulses (up to 7 pulses) for (a) PTFE, (b) nylon, and (c) Al. The temperature profile becomes more non-Gaussian in shape and will end up a linear line if the system size is finite, as shown in Fig. S1(b).

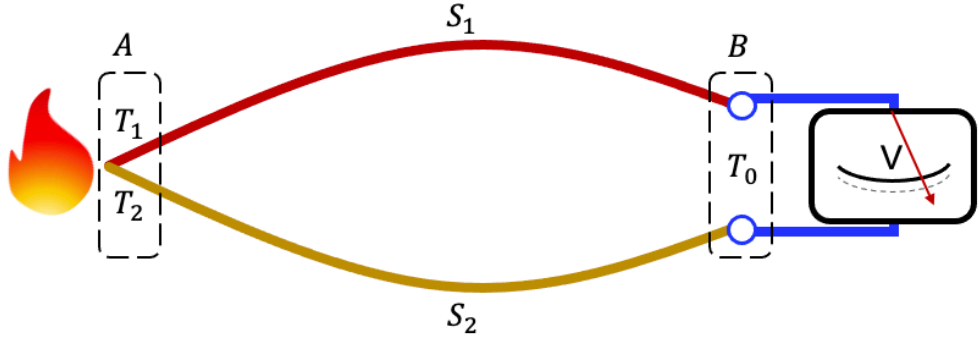

FIG. S4. Voltage in the thermocouple. Schematic of thermocouple used to measure temperature $T_1$ and $T_2$ at the interface (A) between two contacting materials 1 and 2 by measuring the potential difference at (B) with a voltmeter at ambient temperature $T_0$. In a thermocouple, it is generally assumed that two terminal temperatures $T_1$ and $T_2$ are the same, $\Delta T_1 = \Delta T_2$. Then, the voltage difference is expressed as $V_1 - V_2 = S_2\Delta T_2 - S_1\Delta T_1$, equivalent to our triboelectric voltage difference in the main text.

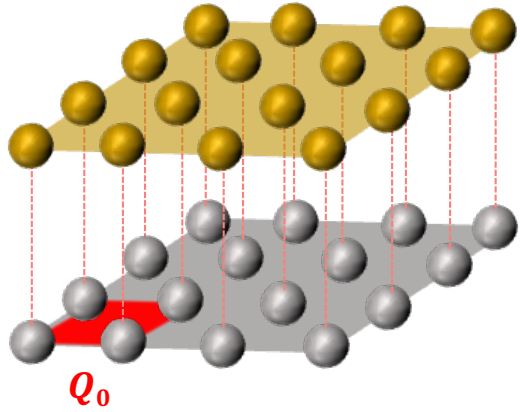

FIG. S5. van der Waals contact. Two ideal materials interact weakly *via* van der Waals interaction, of which strength is about 10 meV per Å². Then, the frictional heat $Q_0$ is assumed to be 1 meV/Å² or 0.01 J/m² per every $\tau = 1$ ns.

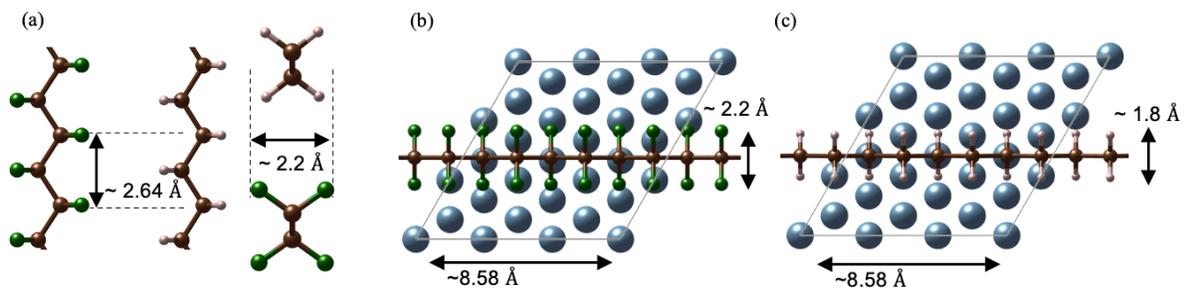

FIG. S6. Atomic structures of (a) PTFE-PE, (b) PTFE-Al, and (c) PE-Al used for van der Waals binding energy calculations.

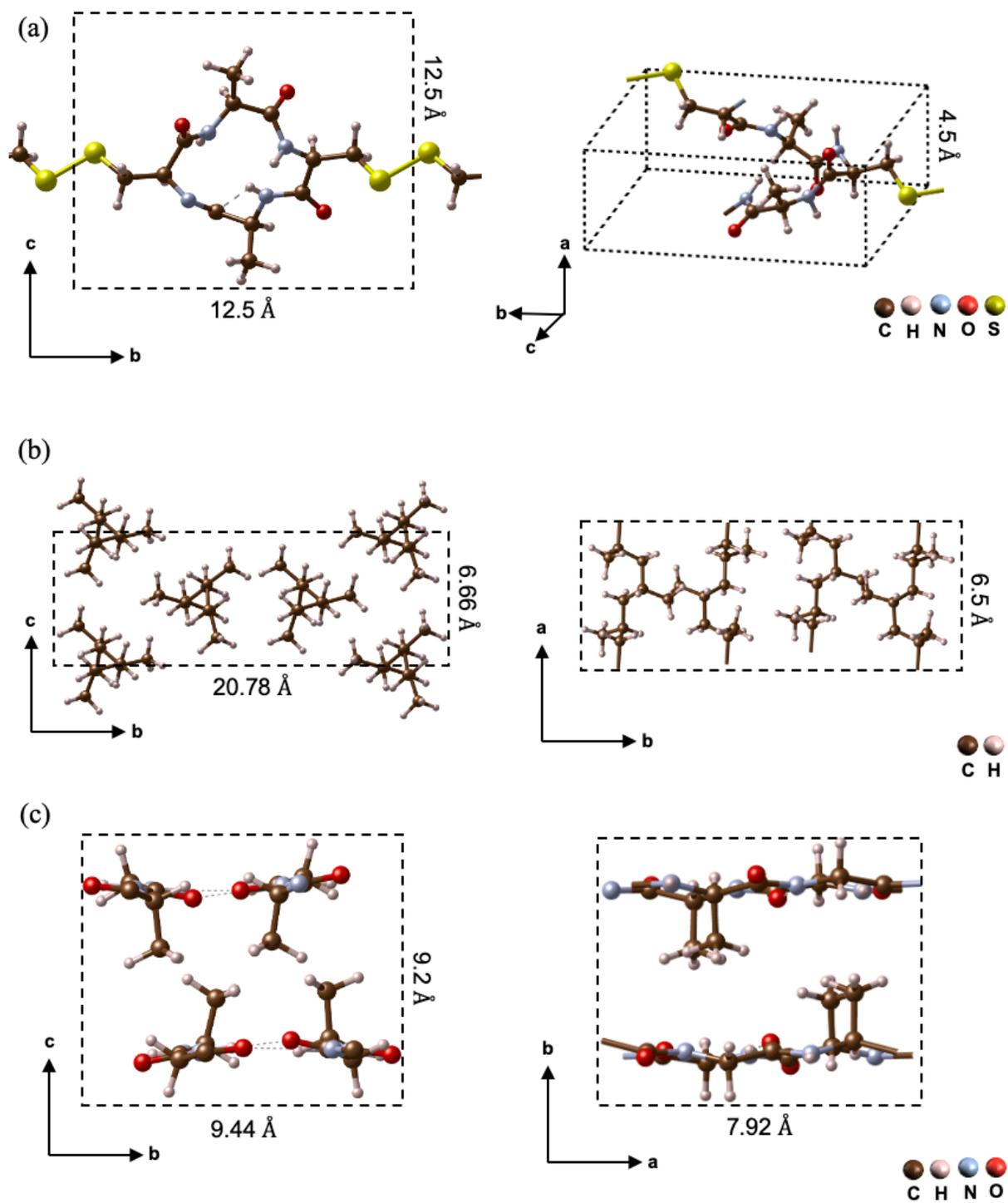

FIG. S7. Atomic structure of triboelectric materials. (a) wool (sulfur cross-linked *α*-Keratin), (b) Polypropylene (PP), and (c) silk (fibroin). The dashed rectangle represents the unit cell. Lattice vectors and constants are also displayed.

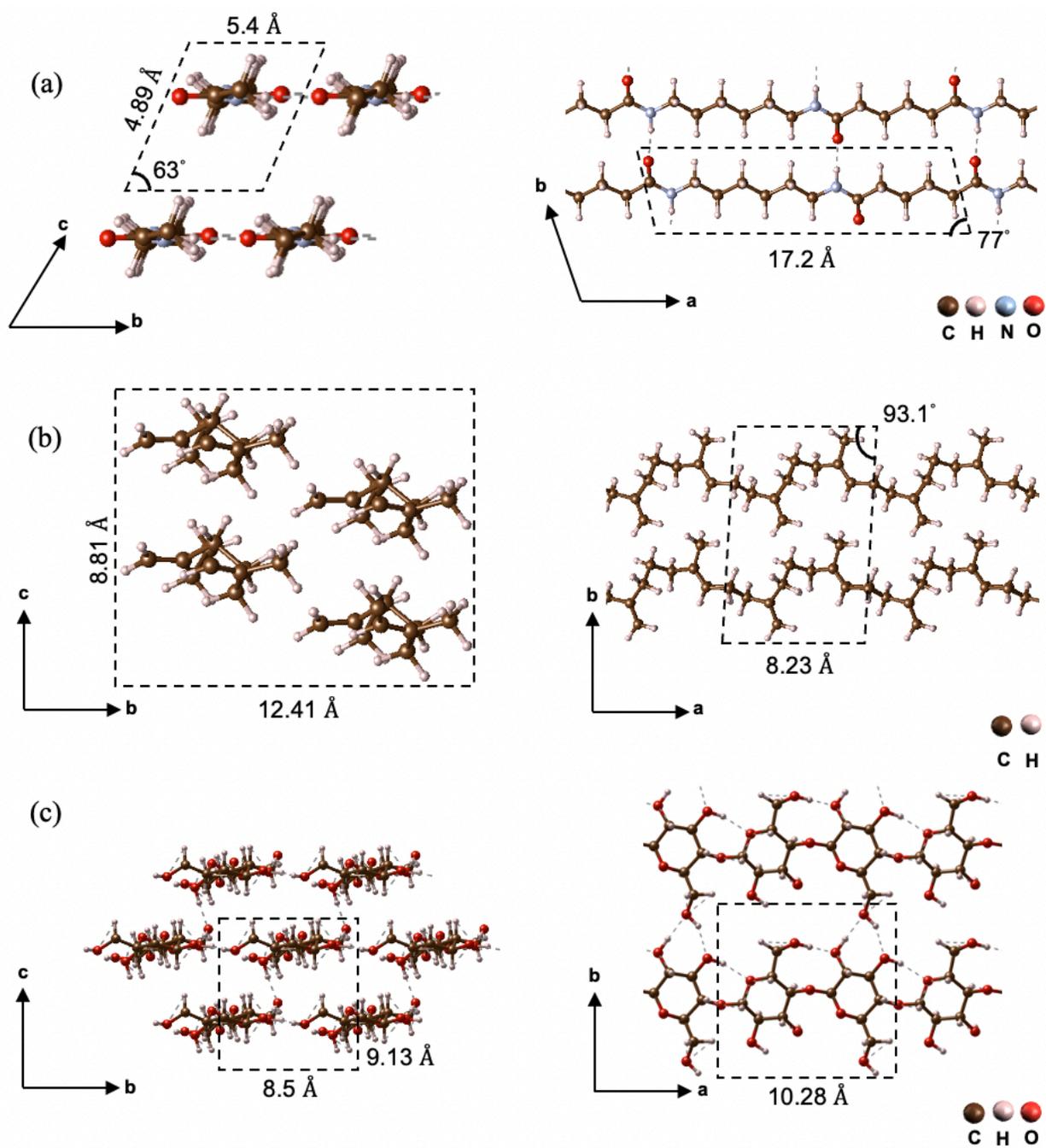

FIG. S8. Atomic structure of triboelectric materials. (a) nylon (nylon66), (b) Polyisoprene (NR: natural rubber), and (c) *Iβ*-Cellulose. The dashed rectangle represents the unit cell. Lattice vectors and constants are also displayed.

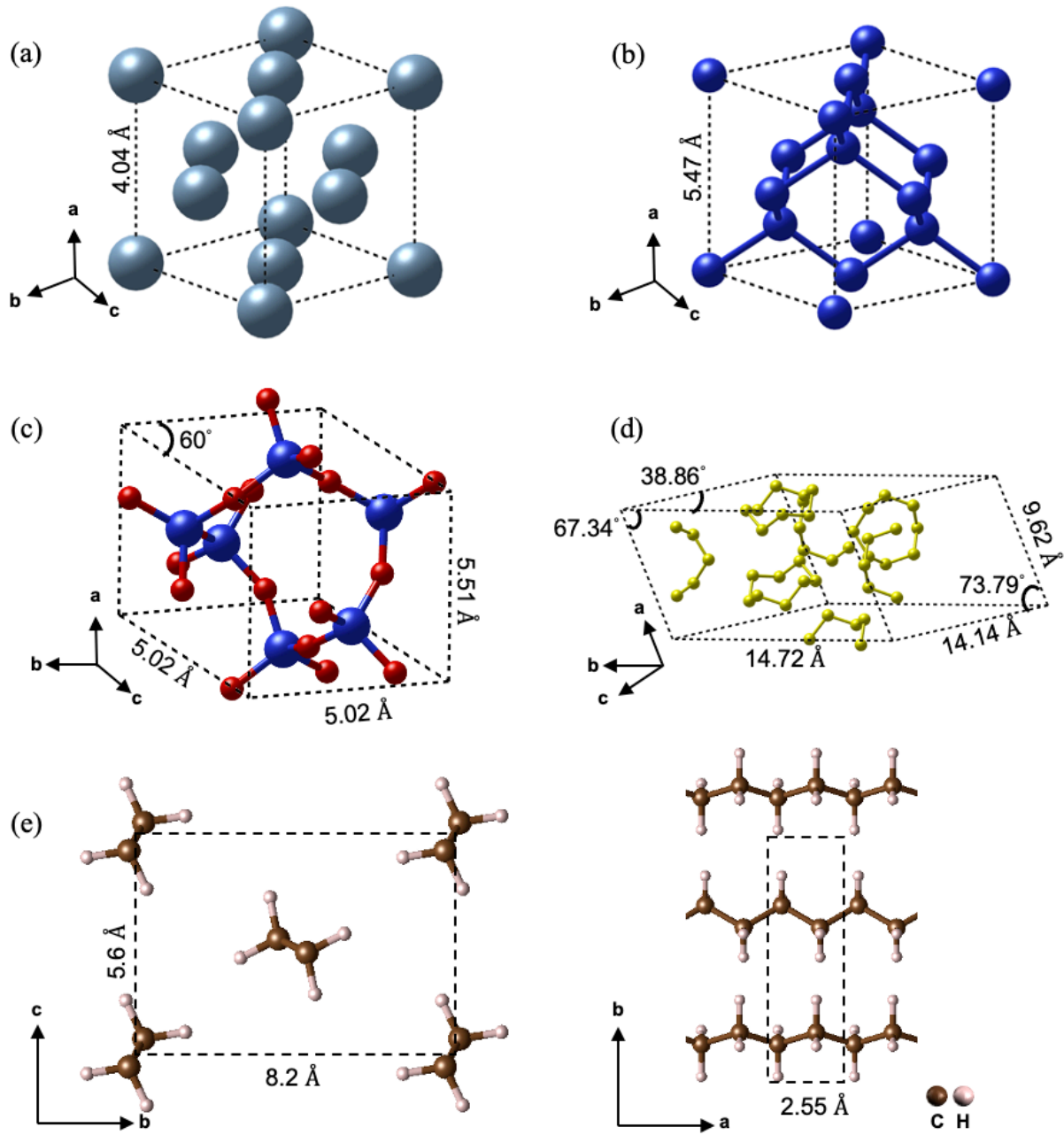

FIG. S9. Atomic structure of triboelectric materials. (a) Al, (b) Si, (c) quartz (Trigonal SiO2), (d) Orthorhombic *α*-sulfur, and (e) Polyethylene (PE). The dashed rectangle represents the unit cell. Lattice vectors and constants are also displayed.

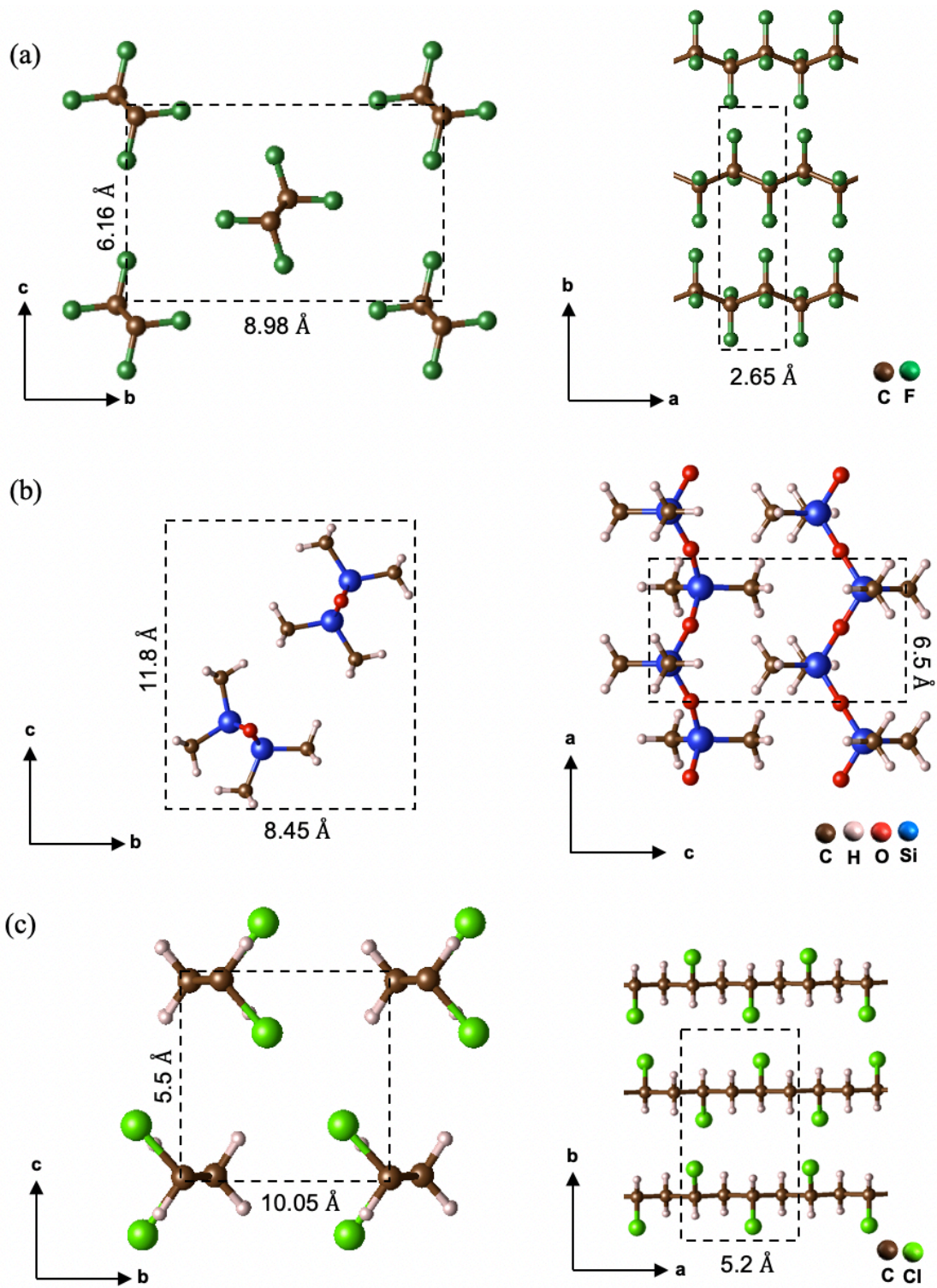

FIG. S10. Atomic structure of triboelectric materials. (a) Polytetrafluoroethylene (PTFE), (b) Polydimethylsiloxane (PDMS: silicone rubber), (c) Polyvinyl chloride (PVC). The dashed rectangle represents the unit cell. Lattice vectors and constants are also displayed.

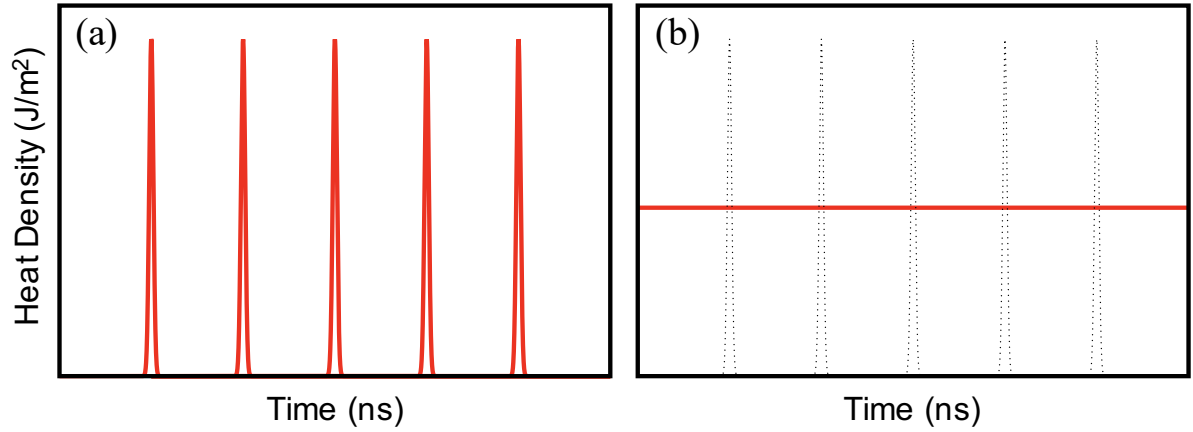

FIG. S11. Heat pulses *vs.* continuous heat. The frictional heat is generated (a) in the form of successive pulses or (b) continuously. For the continuous heat injection, the heat source can be expressed as $\dot{Q}(x,t) = \frac{1}{2}\frac{Q_0}{\tau}\delta(x)$.

TABLE S1. Heat partition between two contacting materials described in Fig. 3 and its corresponding temperature profile for a single heat pulse at the interface. Indexes 1 and 2 indicate the two materials. No temperature gap appears in the thermal diffusion length scheme.

| | Heat partition | Temperature profile |
|---|---|---|
| First atomic layer | $\dfrac{\rho_{1,2} c_{1,2}}{\rho_1 c_1 + \rho_2 c_2}$ | $\dfrac{Q_0}{\rho_1 c_1 + \rho_2 c_2} \dfrac{e^{-\frac{x^2}{4\alpha_{1,2} t}}}{\sqrt{\pi \alpha_{1,2} t}}$ |
| Phonon mean-free-path | $\dfrac{\rho_{1,2} c_{1,2} v_{1,2}}{\rho_1 c_1 v_1 + \rho_2 c_2 v_2}$ | $\dfrac{Q_0 v_{1,2}}{\rho_1 c_1 v_1 + \rho_2 c_2 v_2} \dfrac{e^{-\frac{x^2}{4\alpha_{1,2} t}}}{\sqrt{\pi \alpha_{1,2} t}}$ |
| Thermal diffusion length | $\dfrac{\sqrt{\rho_{1,2} c_{1,2} k_{1,2}}}{\sqrt{\rho_1 c_1 k_1} + \sqrt{\rho_2 c_2 k_2}}$ | $\dfrac{Q_0}{\sqrt{\rho_1 c_1 k_1} + \sqrt{\rho_2 c_2 k_2}} \dfrac{e^{-\frac{x^2}{4\alpha_{1,2} t}}}{\sqrt{\pi t}}$ |
| Half and half | $\dfrac{1}{2}$ | $\dfrac{Q_0}{2} \dfrac{e^{-\frac{x^2}{4\alpha_{1,2} t}}}{\sqrt{\pi \rho_{1,2} c_{1,2} k_{1,2} t}}$ |

TABLE S2. van der Waals binding energy per contact area. The binding energy $E_b$ is defined as $E_{A+B} - E_A - E_B$. Here, $E_A$ and $E_B$ are total energy of systems A and B.

|  | $E_b$ (meV) | $A$ (Å²) | $E_b/A$ (meV/Å²) |
|---|---|---|---|
| PTFE-PE | 41 | 5.8 | 7.0 |
| PTFE-Al | 220 | 19 | 12 |
| PE-Al | 380 | 15 | 20 |

TABLE S3. Calculated macroscopic dielectric constants ($\varepsilon_{PBE}^{\infty}$) and HSE06 mixing parameters ($\alpha_{HSE06}$) are listed for various triboelectric materials.

|           | $\varepsilon_{PBE}^{\infty}$ | $\alpha_{HSE06}$ |
|-----------|------|------|
| wool      | 1.85 | 0.54 |
| PP        | 2.21 | 2.31 |
| silk      | 0.45 | 0.43 |
| nylon     | 2.04 | 0.49 |
| NR        | 2.41 | 0.42 |
| cellulose | 2.17 | 0.46 |
| Al        | 46   |      |
| Si        |      | 0.25 |
| quartz    | 2.32 | 0.43 |
| sulfur    | 2.63 | 0.38 |
| PE        | 2.05 | 0.49 |
| PTFE      | 1.87 | 0.54 |
| PDMS      | 2.44 | 0.41 |
| PVC       | 2.50 | 0.40 |